\documentclass{aa}
\usepackage{float}
\usepackage{graphicx}
\usepackage{txfonts}
\usepackage{subcaption}
\usepackage{lscape}
\usepackage{placeins}
\usepackage[colorlinks=true,linkcolor=blue,allcolors=blue]{hyperref}
\usepackage{pifont}
\newcommand{\cmark}{\ding{51}}%
\newcommand{\xmark}{\ding{55}}%

\newcommand{\lya}{Ly$\alpha$}
\newcommand{\ha}{H$\alpha$}
\newcommand{\hb}{H$\beta$}

\newcommand{\civ}{\ion{C}{iv}}
\newcommand{\ciii}{\ion{C}{iii}]}
\newcommand{\heii}{\ion{He}{ii}}
\newcommand{\fesc}{f_\mathrm{esc}}

\newcommand{\xion}{\xi_\mathrm{ion}}

\newcommand{\kms}{\,\mathrm{km\,s^{-1}}}
\newcommand{\DJA}{\textsc{DJA}}

\defcitealias{naidu2024}{Naidu \& Matthee et al. 2024} 
\begin{document}

 \title{The ionizing properties of JWST's compact broad-line emitters}
 \author{Sara~Mascia\inst{\ref{inst:ista}}\corrauth{sara.mascia@ista.ac.at}
  \and Jorryt~Matthee\inst{\ref{inst:ista}}
  \and Alberto~Torralba\inst{\ref{inst:ista}}
  \and 
  Jenny~E.~Greene\inst{\ref{inst:Princeton}}
  \and Anna-Christina~Eilers\inst{\ref{inst:MIT},\ref{inst:MIT_kavli}}
  \and Edoardo~Iani\inst{\ref{inst:ista}}
  \and Rohan~P.~Naidu\inst{\ref{inst:MIT_kavli}, \ref{inst:hawaii}}
  }

 \institute{Institute of Science and Technology Austria (ISTA),
    Am Campus 1, A-3400 Klosterneuburg, Austria\label{inst:ista}
    \and Department of Astrophysical Sciences, Princeton University, 4 Ivy Lane, Princeton, NJ08544, USA\label{inst:Princeton}
    \and Department of Physics, Massachusetts Institute of Technology, Cambridge, MA 02139, USA \label{inst:MIT}
    \and MIT Kavli Institute for Astrophysics and Space Research, Massachusetts Institute of Technology, Cambridge, MA 02139, USA\label{inst:MIT_kavli}
    \and Institute for Astronomy, University of Hawai'i, 2680 Woodlawn Drive, Honolulu, HI 96822, USA \label{inst:hawaii}}

 \date{Accepted XXX. Received YYY; in original form ZZZ}

 \abstract
 {The recent JWST discovery of a numerous population of broad-line emitters (BLEs) at $z > 4$ has reopened the question of whether sources other than faint galaxies contributed significantly to cosmic reionization. To assess this contribution, we present a systematic census of compact, blue broad-line emitters at $4 \leq z \leq 7$ selected from the Dawn JWST Archive (DJA), designed to isolate sources in which the UV continuum and the broad-line emission originate from the same compact physical region. Starting from a parent sample of $4145$ galaxies with NIRSpec/PRISM spectroscopy and NIRCam imaging, we apply criteria on broad \ha\ emission ($\mathrm{FWHM} > 2000$\,km\,s$^{-1}$ detected at $\mathrm{S/N}\geq5$, complete at $L_{\rm H\alpha,broad}\gtrsim10^{42.7}$\,erg\,s$^{-1}$), blue continuum slopes ($\beta_\mathrm{UV} \leq -1.5$, $\beta_\mathrm{opt} < 0.5$), and morphological compactness in both the rest-frame UV and optical, returning a final sample of 20 sources. This constitutes $\sim 20$\% of the PRISM-selected broad-line sample that is dominated by Little Red Dots (LRDs).
 We find that the compact blue BLEs display hot ionizing continua, high-ionization UV lines (\ion{C}{iv}, \ion{N}{iv]}, \ion{He}{ii}), and elevated \lya\ emitter fractions ($X_{\rm Ly\alpha} = 50$--$67$\%) relative to star-forming galaxies and LRDs. Beyond their similar lack of X-ray emission, their optical spectra closely resemble those of LRDs. This suggests that compact blue BLEs extend this population towards lower column density envelopes, but with similar engines, rather than representing a physically distinct population. Using \textsc{Sirocco} to model physical configurations that reproduce the observed spectra, we find a high ionizing photon production efficiency ($\log \xi_\mathrm{ion}/{\rm Hz\,erg^{-1}} \sim 25.4$) and a weighted average escape fraction $f_\mathrm{esc} \sim 0.2$ for the bluest sources, and $\sim 0.01$ for LRDs. This implies that broad-line sources do not dominate reionization globally, but given their luminosity, their contribution can dominate up to few Mpc scales.}

 \keywords{galaxies: active -- galaxies: high-redshift --
    galaxies: evolution -- galaxies: nuclei --
    dark ages, reionization}

 \maketitle

\section{Introduction}
\label{sec:intro}
The epoch of reionization (EoR) marks the last major phase transition of baryonic matter in the Universe, during which the neutral hydrogen of the intergalactic medium (IGM) was ionized by early galaxies \citep[for a recent review, see][]{Stark2026}.
The timeline of reionization is relatively well constrained: hydrogen reionization is largely complete by $z \sim 5.3$--$5.5$ as inferred from the statistics of the \lya\ forest measured in quasars \citep[e.g.,][]{Bosman2022} and the optical depth to Thomson scattering measured in the cosmic microwave background constrains the average redshift at $z\approx7.7$ \citep{collaboration2020}. 

The canonical picture is that faint star-forming galaxies (SFGs) are the primary contributors to the reionization photon budget, based on the fact that their comoving number density is much higher than that of active galactic nuclei (AGN) during the EoR, and that even modest Lyman Continuum (LyC) escape fractions ($f_{\rm esc}^{\rm LyC}\approx10$\%) are sufficient to sustain reionization when integrated over the faint end of the UV luminosity function \citep[e.g.,][]{Robertson2015, Finkelstein2019, Atek2026, Simmonds2024, Giovinazzo2026}. 
Observational support for this picture has grown substantially in recent years, with direct detections of LyC emission from $z \sim 3$--$4$ and $z \sim 0.3$ SFGs \citep[e.g.,][]{Riverathorsen17, Steidel2018, Izotov2018, Flury2022, Gupta2024} that have enabled estimates of $f_{\rm esc}^{\rm LyC}$ of relatively bright galaxies in the EoR \citep[$M_{\rm UV} \leq -18$; e.g.,][]{Mascia24} based on indirect indicators that are in line with the required values. Complementary evidence has emerged from cross-correlation measurements between galaxies and the \lya\ forest transmission at the closing of reionization, which provide a statistical, IGM-averaged estimate of the LyC emissivity of the galaxy population as a whole \citep{Kashino2023, Kashino2026, Kakiichi2025}. 
Nevertheless, the averaged escape fraction of the full galaxy population remains uncertain because virtually no direct constraints exist on the $f_{\rm esc}^{\rm LyC}$ of galaxies fainter than M$_{\rm UV}>-19$.  These account for the vast majority of the galaxy population at these redshifts, contributing an estimated $\gtrsim 75$\% of the total UV luminosity density at $z \sim 6$ \citep{Atek2026}, and their collective ionizing output is therefore the most poorly anchored component of any reionization budget. Moreover, the role of rare, bright starbursts with potential significant leakage \citep[e.g.,][]{MarquesChaves26a} remains unclear \citep[e.g.,][]{Sharma2017, Naidu2018,Matthee2022}. 

Recent JWST observations have added a new layer of complexity to this picture. The detection of \lya\ emission at $z > 10$ in several sources \citep{Witstok2025, Bunker2023} and the apparent weak damping wing in MoM-z14 at $z=14.4$ \citep{Naidu2025} are difficult to reconcile with the high neutral hydrogen fraction expected at these redshifts, pointing to local ionization by exceptionally luminous sources. Do such rare, compact sources that may be capable of carving out large ionized bubbles in an otherwise predominantly neutral IGM play a disproportionate role in reionization?

AGN have been considered subdominant contributors to reionization primarily because the space density of quasars declines sharply above $z \sim 3$ \citep{Hopkins2007, Kulkarni2019}. However, AGN may play a more significant role than previously assumed. A faint AGN population had been proposed as a significant driver of reionization on the basis of pre-JWST X-ray and UV constraints \citep{Madau2015, Giallongo2019}, but observational support for such a population remained limited. Deep spectroscopic surveys with JWST have revealed a numerous population of moderate luminosity, broad-line emitters at $4 \lesssim z \lesssim 9$ \citep[$M_{\rm UV} \approx -18$ to $-21$; e.g.,][]{Kocevski2023,Harikane2023, Greene2024, Matthee2024, Maiolino_JADES_2024, Juodzbalis2025}, accounting for $\sim1\%$ among UV-selected galaxies at $z \sim 4-7$. These number densities have prompted a renewed reassessment of AGN-dominated reionization scenarios. In particular, \citet{Madau2024} showed that, under certain assumptions on the AGN fraction, escape fraction \citep{Cristiani2016, Grazian2018, Romano2019}, and UV spectral slope, a population of faint Type-1 AGN can reproduce both the hydrogen and helium reionization history and the Thomson scattering optical depth.

However, a key challenge is that the broad-line emitter population identified by JWST is far from homogeneous. A substantial fraction consists of the so-called \textit{Little Red Dots} \citep[LRDs;][]{Matthee2024}, characterized by blue UV and red optical colors, compact optical morphology and broad Balmer lines \citep{Kocevski2023, Greene2024, Kokorev2024, Matthee2024, Labbe2025}. These are interpreted as AGN highly covered by dense gas \citep[e.g.,][]{Torralba2026a, Naidu2025, Inayoshi2025, deGraaff2025b} from which few UV photons escape \citep[but see][for indications of holes in the envelope through which some UV emission may escape]{Tang26, Ji2026, Torralba2026}. Potentially more relevant for reionization is that a fraction of broad-line emitters display blue UV continua and no Balmer absorption characteristic of a dense, high-column gas envelope \citep{Ubler2023, Brazzini2026, Geris2026, Matthee2026}. However, the fraction of the galaxy population that comprises such sources has not been systematically explored yet. Whether the newly identified broad-line source contributed to reionization depends on their $f_{\rm esc}^{\rm LyC}$: direct measurements are precluded at $z \gtrsim 4$ by IGM absorption \citep{Inoue2014}, and indirect constraints from analogous sources at lower redshift, where the LyC is observable, do not yet exist either.

An alternative approach to measuring $f_{\rm esc}^{\rm LyC}$ consists of modeling the observed spectra of these sources under assumptions on their geometry and gas distribution \citep{Sneppen2026a, Gentile2026}. Such models have reproduced key observables of LRDs including the strength of the Balmer break and the \ha\ line profile. Yet, no study has used this framework to estimate the ionizing output from these sources.

In this paper, we take a step toward filling this gap by presenting a systematic study of blue, compact broad-line emitters at $4 \leq z \leq 7$ drawn from the Dawn JWST Archive \citep[DJA;][]{Heintz2023, deGraaff2025}, selected to have simultaneous coverage of \lya\ and \ha\ with JWST/NIRSpec.  
Our analysis addresses four key questions:
\textit{i.} What fraction of sources exhibit bright, blue UV emission that is predominantly powered by the same region as the broad lines?
\textit{ii.} What are the UV continuum and emission line properties of the blue, compact broad-line emitters as a population?
\textit{iii.} What spectral diagnostics provide the most effective constraints on the nature and hardness of the ionizing continuum, and can AGN accretion and stellar processes be distinguished using the available data?
\textit{iv.} What is the contribution of this population to the reionization photon budget and how sensitive is this estimate to the assumed powering mechanism, surrounding gas geometry, and structure?

The paper is structured as follows. We describe the data and observational setup in Sect.~\ref{sec:data}. The sample selection criteria are detailed in Sect.~\ref{sec:sample}. The spectral characterization of the sample is presented in Sect.~\ref{sec:spectra}. We present photoionization modeling with \textsc{Sirocco} in Sect.~\ref{sec:sirocco}. The contribution to reionization is quantified in Sect.~\ref{sec:discussion}, where we also discuss the implications of our results. We summarize our conclusions in Sect.~\ref{sec:conclusions}. Throughout this paper, we assume a flat $\Lambda$CDM cosmology with $H_0 = 70\,\mathrm{km\,s^{-1}\,Mpc^{-1}}$, $\Omega_\mathrm{M} = 0.3$, and $\Omega_\Lambda = 0.7$.


\section{Data}
\label{sec:data}

\subsection{NIRSpec}
\label{subsec:nirspec}

Our spectroscopic analysis is based on spectra obtained with the NIRSpec multi-shutter assembly (MSA) on JWST. We use observations compiled in version 4.5 of the DJA\footnote{\url{https://dawn-cph.github.io/dja/}}. This dataset includes major public surveys such as CANUCS \citep{Sarrouh2025}, CAPERS (GO-6368; PI: Dickinson), CEERS \citep{Finkelstein2025}, JADES \citep{Eisenstein2023, CurtisLake2025, Scholtz2025}, NEXUS \citep{Shen2024}, NIRSpec GTO-Wide \citep{Maseda2024}, RUBIES \citep{deGraaff2025}, UNCOVER \citep{Bezanson2024, Price2025}, and GO-8204 (PIs Greene and Labb\'e) and several  smaller programs. We also included observations from NIRSpec IFU programs, GA-NIFS (GO-1216, PI Lützgendorf), and GO-5664 (PI: Matthee). All spectra were reduced uniformly using the \texttt{msaexp} software package \citep{Brammer2023}, implementing techniques consistent with recent analysis of deep NIRSpec datasets \citep{Heintz2023, deGraaff2025}. Where feasible, we employed local background subtraction utilizing nodded exposure sequences to minimize contamination from sky emission and neighboring sources. 

Version 4 of the DJA implements empirical wavelength calibration corrections based on source centroid positions within MSA shutters, addressing systematic offsets between PRISM and higher-resolution modes \citep{deGraaff2025b}, and incorporates updated flux calibration reference files that extend extraction beyond nominal wavelength ranges \citep{Valentino2025, Pollock2025}.

1D spectra were extracted using an optimal extraction algorithm \citep{Horne1986}, which weights pixels according to the light profile of the object in the spectrum to maximize the signal to noise. The resulting extraction kernel, combined with precise source positioning within each shutter, enabled wavelength-dependent slit loss corrections which, for point-like sources, have been demonstrated to perform reliably \citep[e.g.,][]{deGraaff2025,Hviding2025} and were therefore applied without additional flux adjustments. All reduced spectra underwent comprehensive visual inspection to verify spectral quality and validate automated redshift determinations \citep[see][for further details]{deGraaff2025b}. We selected only sources with grade = 3 redshifts \citep[robust redshifts,][]{deGraaff2025b}.

For sources observed in gravitationally lensed fields, we applied magnification corrections to all derived luminosity measurements. Specifically, we adopted the Abell 2744 lens model from \citet{Furtak2023a} and \citet{Price2025} for UNCOVER sources, while magnification estimates in CANUCS cluster fields were released in \cite{Sarrouh2025}.

\subsection{NIRCam}\label{subsec:nircam}

Deep, high resolution imaging data in the near-infrared were obtained from JWST/NIRCam mosaics compiled in version 7 of the DJA, constructed at a pixel scale of 0.04$''$. Reductions were performed using the \texttt{grizli} pipeline \citep{Brammer2023b}, with detailed processing methodology documented in \citet{Valentino2023}. For this work, we focus on both rest-frame UV and rest-frame optical morphological properties of our sources. We extracted imaging data in all available NIRCam filters from the aforementioned JWST programs. This includes the following filters where available: F070W, F090W, F115W, F150W, F200W, F277W, F356W, and F444W. 

We note that temporal offsets between spectroscopic observations and subsequent NIRCam imaging campaigns can introduce minor astrometric discrepancies. This is particularly relevant for early NIRSpec programs that relied on HST-based target selection prior to the availability of JWST imaging. To mitigate potential systematic offsets, we cross-matched our spectroscopic catalog against comprehensive photometric catalogs maintained on the DJA platform, obtaining refined source centroid positions based on NIRCam detections. Photometric measurements were then extracted from image cutouts centered on these cross-matched positions using circular apertures with radii of $0.1\arcsec$ and $0.2\arcsec$, consistent with standard approaches for compact source photometry \citep[e.g,][]{Kokorev2024, Labbe2025}.


\section{Sample selection}
\label{sec:sample}

Our parent sample consists of 4145 sources drawn from the \DJA\ spectroscopic compilation described in Sect.~\ref{sec:data}, selected to have simultaneous rest-frame spectral coverage of \lya\ to \ha\ at $4 \leq z \leq 7$. To isolate compact, blue sources with a significant broad \ha\ emission component, we apply three sequential selection criteria which we describe in detail below. The number of sources retained at each step is summarized in Table~\ref{tab:selection}.

\subsection{\texorpdfstring{Broad H$\alpha$ emission}{Broad Ha emission}}
\label{sec:ha_selection}

The first step is to select sources with strong, broad \ha\ emission, regardless of their broad-band colors. To maximize our statistics we use PRISM spectra to identify these broad lines, which implies that our sample is limited to sources with significant contribution from broad lines.

We fit the \ha\ emission line of each source in the parent sample with two competing models: a single Gaussian plus continuum, and a two-component narrow-plus-broad Gaussians plus continuum. The continuum is in both cases modeled as a power-law fitted locally over two sidebands on either side of \ha, over rest-frame windows of 6150--6450\,\AA\ and 6650--6750\,\AA. Both models are fitted to the PRISM spectra using a Levenberg--Marquardt least-squares minimization. The instrumental line spread function of NIRSpec/PRISM is approximated as a wavelength-dependent Gaussian with $\sigma_\mathrm{LSF}$ computed from the point-source resolution curve, and convolved in quadrature with the intrinsic line profile. Parameter uncertainties are propagated via Monte Carlo sampling of the best-fit parameter covariance matrix, drawing $N_\mathrm{MC} = 1000$ realizations per source.

\begin{figure}
\centering
\includegraphics[width=0.92\linewidth]{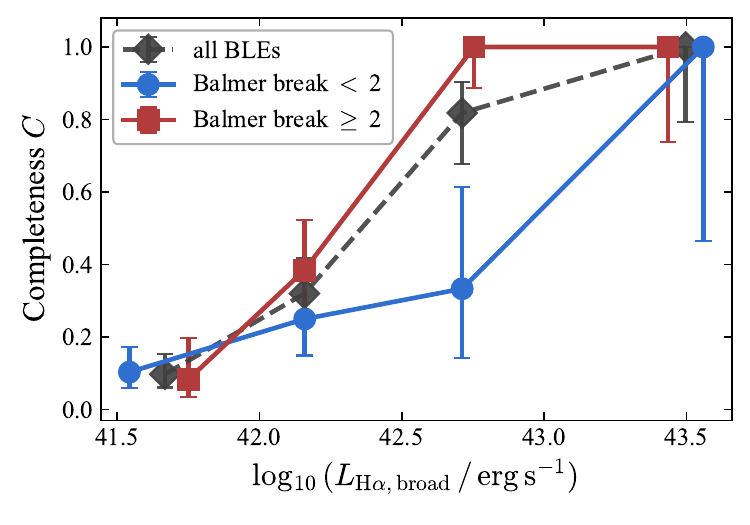}
\caption{The completeness of detecting broad lines in PRISM data as a function of broad-\ha\ luminosity $L_{H\alpha \rm ,broad}$. The recovered fraction of grating-confirmed broad-line emitters is shown for all sources (black, dashed), for the population with weak ($<2$) Balmer breaks (blue circles), and for the sources with Balmer breaks $\geq$ 2 (red squares). Points mark the median luminosity of each bin; error bars are $68\%$ binomial confidence intervals.}
\label{fig:completeness}
\end{figure}

Broad-line emitters (BLEs) are identified following the three sequential criteria of \cite{Greene2024}. First, the improvement in $\chi^2$ between the two models must satisfy $\Delta\chi^2 > 11.5,$ (C1), corresponding to the $3\sigma$ significance threshold for four additional degrees of freedom. Second, the FWHM of the broad component must exceed $2000\,\kms$ (C2). Third, the signal-to-noise (S/N) ratio of the broad component, must satisfy $\mathrm{S/N}_\mathrm{b} \geq 5$ (C3). Sources satisfying C1 and C2 but failing C3 are flagged as candidate BLEs and excluded from subsequent analysis. Only sources satisfying all three criteria simultaneously are retained as confirmed broad-line emitters. All the PRISM-selected sources with usable grating observations are confirmed BLEs (100\% purity), i.e. broad \ha\ is detected with S/N $\geq 3$ in both dispersers. However, our PRISM selection recovers only the 30\% of the grating-selected BLEs. The incompleteness depends strongly on both the \ha\ luminosity and line width (Fig.~\ref{fig:completeness}): the recovery rate of grating‑confirmed BLEs in the PRISM rises from 11\% at $L_{H\alpha \rm ,broad} < 10^{42} \ \text{erg s}^{-1}$ to 100\% above $10^{43}$ $\text{erg s}^{-1}$, and because the PRISM resolution is redshift-dependent, from 18\% at $z<5$ to 67\% at $z>6$. These trends are not independent: more luminous broad-\ha\ lines are also intrinsically broader, and both properties make the broad component easier to detect at PRISM resolution. The sources we preferentially recover are therefore the luminous, broad-lined emitters, which are also those with the strongest Balmer breaks (Fig.~\ref{fig:completeness}). Overall, our selection reaches 50\% completeness at $L_{\rm H\alpha,broad}\approx10^{42.3}$\,erg\,s$^{-1}$ and 90\% completeness at $\approx10^{42.7}$\,erg\,s$^{-1}$, and is substantially incomplete below these luminosities, where we miss the majority of the fainter broad-\ha\ population.

\subsection{UV and optical continuum slopes}
\label{sec:beta_selection}

The UV and optical continuum slopes are measured by fitting a power law $F_\lambda \propto \lambda^{\beta}$ to the rest-frame PRISM spectra over two wavelength windows: $1300$--$3000$\,\AA\ for the UV slope $\beta_\mathrm{UV}$, and $5000$--$6500$\,\AA\ for the optical slope $\beta_\mathrm{opt}$. The regression is performed by weighted least-squares in log--log space, with known emission lines masked within $\pm100$\,\AA\ of their rest-frame wavelengths (\lya, \ion{C}{IV}, \ion{He}{II}, \ion{O}{III}], \ion{C}{III}], \ion{Mg}{II} in the UV window; \hb, [\ion{O}{III}], \ha, [\ion{N}{II}] in the optical window; at PRISM resolution several of these are blended),  leaving sufficient line-free continuum for a robust power-law fit in both windows.  We verified that the resulting slopes are consistent with those derived independently from broadband NIRCam photometry (F090W--F200W for $\beta_\mathrm{UV}$; F277W--F444W for $\beta_\mathrm{opt}$) for sources where both measurements are available. We additionally measure the Balmer break strength as the ratio of the median flux density in the $5400$--$5700$\,\AA\ window to that in the $3500$--$3650$\,\AA\ window \citep{Wang2025,deGraaff2025a}. 

\cite{Kocevski2025} select LRD candidates with $-2.8 < \beta_\mathrm{UV} < -0.37$, where the lower limit excludes brown dwarf contaminants and the upper limit avoids dust-reddened systems \citep[][]{Brazzini2026, Geris2026}. Our broad-\ha\ selection (Sect.~\ref{sec:ha_selection}) already removes the possible contaminants from the sample, so the role of the $\beta_\mathrm{UV}$ cut is not to identify AGN but to isolate sources in which the UV continuum reaches the observer with minimal attenuation. We therefore adopt a more stringent threshold of $\beta_\mathrm{UV} \leq -1.5$, motivated by the empirical observed slope of unreddened quasar composites \citep{VandenBerk2001}: any source redder than $\beta_\mathrm{UV} \sim -1.5$ is unlikely to be observed without significant dust attenuation along the line of sight, and its UV slope would no longer reliably trace the intrinsic continuum of the central source. Of the 99 confirmed broad-line emitters, 35 have $\beta_\mathrm{UV} > -1.5$; these are predominantly LRDs (31), whose redder UV slopes reflect attenuation by the dense gas envelope, and a small number of classical broad-line AGN (4) reddened by dust along the line of sight.
We further require $\beta_\mathrm{opt} < 0.5$ to exclude the reddest sources in the optical. Based on a visual inspection of their spectra, sources above this threshold in our sample are either obscured quasars (2) or LRDs (25) not compact in the UV, therefore not representative of the compact blue BLE population we aim to characterize. The measured slopes and Balmer break strengths for the final sample are reported in Table~\ref{tab:sample}. Fig.~\ref{fig:beta_diagram} shows the distribution of $\beta_\mathrm{UV}$ versus $\beta_\mathrm{opt}$ for the compact blue BLEs, the other BLE sample, and the SFG comparison sample.

\subsection{Morphological compactness}
\label{sec:compact_selection}

Morphological compactness has been widely used to identify high-redshift AGN candidates and LRDs, with most studies relying on a single compactness criterion measured in the rest-frame optical, typically the F444W band \citep[e.g.,][]{Kocevski2025,Hviding2025,Kokorev2024}. However, a compact optical morphology alone does not guarantee equally compact rest-frame UV emission, which can be contaminated by an extended host galaxy and bias our study of the UV emission from the broad-line region \citep{Chen24,Rinaldi25,Baggen2026, Ishikawa2026}. To mitigate this contamination, require compactness in both the rest-frame UV and optical, ensuring that the blue continuum genuinely originates from a point-like source. 

Source compactness is assessed independently in the rest-frame UV and rest-frame optical using the NIRCam imaging described in Sect.~\ref{subsec:nircam}. We quantify compactness via the ratio of fluxes enclosed within circular apertures of $0.2\arcsec$ and $0.1\arcsec$ radius centered on the source position, $\mathcal{C} \equiv \frac{F_{0.2\arcsec}}{F_{0.1\arcsec}}$,
measured on the NIRCam mosaics using \textsc{photutils} \citep{Bradley2024}. No PSF aperture correction is applied to the flux ratio: our aperture sums are extracted directly from the NIRCam mosaic pixel data, so no de-correction step is needed before computing $\mathcal{C}$. Sources are classified as compact if $\mathcal{C} \leq 1.7$ in both the rest-frame UV and optical. In the rest-frame UV, we adopt a filter fallback chain in order of preference: F115W, F150W, F200W, selecting the first available filter for each source. In the rest-frame optical we use F444W. We note that our compactness criterion is more tolerant than a pure point-source criterion to allow for closely separated clumps. In the case of clumpy systems, we verify through visual inspection that there is blue, UV bright, compact emission co-spatial with the broad Balmer line-emission.

Flux uncertainties are estimated from empty-aperture statistics: 200 random apertures of the same radius are placed in source-free regions of each image cutout, excluding a zone of radius $ 0.3\arcsec$ around the source centroid, and the standard deviation of their flux is adopted as $\sigma_\mathrm{flux}$. When the cutout geometry does not permit a sufficient number of valid sky placements, we fall back to a background RMS estimated from sigma-clipped border pixels, scaled by the square root of the aperture area.

\begin{table}
\caption{Sample selection.}
\label{tab:selection}
\centering
\begin{tabular}{lc}
\hline\hline
Selection step & $N$ \\
\hline
Parent sample ($4 \leq z \leq 7$)   & 4145 \\
Confirmed broad-line emitters     & 99 \\
\quad Compact in rest-frame UV ($\mathcal{C}_\mathrm{UV} \leq 1.7$) & 71 \\
\quad Compact in rest-frame optical ($\mathcal{C}_\mathrm{opt} \leq 1.7$) & 65 \\
Compact in UV and optical, $\beta_\mathrm{UV} \leq -1.5$, $\beta_\mathrm{opt} \leq 0.5$  & \textbf{20} \\
\hline
\end{tabular}
\end{table}

\subsection{Final sample}
\label{sec:final_sample}

\begin{figure}
\centering
\includegraphics[width=\linewidth]{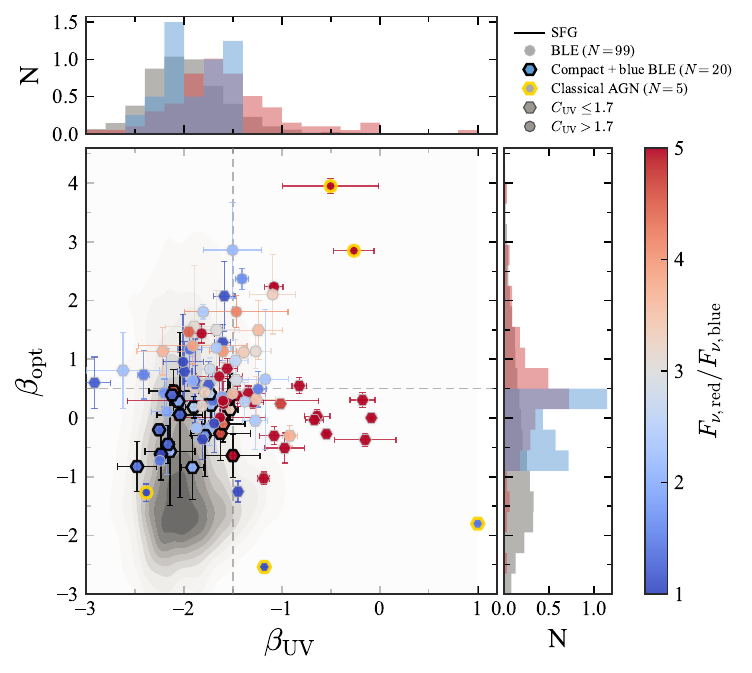}
\caption{Spectroscopic UV versus optical $\beta$ slopes for the population of star-forming galaxies and the broad-line emitters identified in this work. \textit{Main panel}: each BLE is color-coded by Balmer break strength $F_{\nu,\rm red}/F_{\nu,\rm blue}$, shown by the color bar on the right, and the marker shape encodes UV compactness: hexagons indicate sources with $C_{\rm UV} \leq 1.7$, circles $C_{\rm UV} > 1.7$. The 20 compact blue BLEs selected in this work are highlighted with black outlines; sources classified as classical broad-line AGN with yellow outlines. Grey contours show the distribution of DJA star-forming galaxies at $4 \leq z \leq 7$. Dashed lines mark the selection thresholds $\beta_{\rm UV} = -1.5$ and $\beta_{\rm opt} = 0.5$. \textit{Top and right panels}: normalized distributions  of $\beta_{\rm UV}$ and $\beta_{\rm opt}$ for star-forming galaxies (grey),  BLEs (red), and compact blue BLEs (blue). }
\label{fig:beta_diagram}
\end{figure}

Fig.~\ref{fig:beta_diagram} shows the UV and optical colors of the 99 confirmed broad-line emitters, together with the star-forming galaxy (SFG) sample. Broad line sources are color-coded by their Balmer break strength, as this has been widely studied in the context of LRDs \citep[e.g.,][]{Matthee2026,Sneppen2026a}. 

The diagram reveals a broad diversity of spectral types within the BLE population. Sources in the upper-right region ($\beta_{\rm UV} \gtrsim -0.5$, $\beta_{\rm opt} \gtrsim 2$) have red slopes across the UV to optical range, which is indicative of attenuation by dust.
The majority of the 99 PRISM-selected BLEs ($\approx 75$\%) occupy an intermediate region with red optical slopes and moderate-to-strong Balmer breaks whereas their UV slopes are generally blue but probe a relatively wide range ($-2 \lesssim \beta_{\rm UV} \lesssim -1$). By visual inspection and the change in continuum slope, we classify 74 of these sources as LRDs. A small subset of sources, classified as classical broad-line AGN, are highlighted in yellow. Among these, a single source stands out with a very blue optical slope: a luminous quasar observed through the ASPIRE program (GO 2078, PI Wang).
The distribution of SFGs peaks around $\beta_{\rm UV} \approx -2$, $\beta_{\rm opt} \approx -2$, which highlights that the main difference between SFGs and broad-line sources is the optical continuum slope. 

We designed the selection criteria $\beta_{\rm UV} \leq -1.5$ and $\beta_{\rm opt} < 0.5$ (dashed lines in Fig.~\ref{fig:beta_diagram}) to isolate the 20 compact blue BLEs in the lower-left corner of the diagram, where sources are blue in both the UV and the optical and have uniformly weak Balmer breaks ($F_{\nu,\rm red}/F_{\nu,\rm blue} \lesssim 2$ for the majority). These cuts are motivated by the requirement of a hot ionizing radiation field that is uncontaminated by evolved stellar populations or dust reddening, and are complemented by the compactness criteria $\mathcal{C}_{\rm UV}, \mathcal{C}_{\rm opt} \leq 1.7$, which ensure that both the UV and optical continua originate from a single compact physical region rather than from an extended host. Our selection recovers most LRDs from \citet{Geris2026} but few blue broad-line sources, whose faint broad \ha\ falls below the PRISM S/N threshold.

The final sample is likely a lower limit on the true number of compact blue broad-line emitters in the full BLE population, for two reasons. First, in Sec.~\ref{sec:ha_selection} we discuss the incompleteness of our sample: PRISM broad-line selection recovers only the $\sim 30\%$ of the grating-selected broad-line sample, with the incompleteness rising steeply toward low \ha\ luminosity, narrow line width, and low redshift. Blue broad-line emitters are, on average, fainter in broad \ha\ and narrower-lined than LRDs, they are recovered less efficiently (Fig.~\ref{fig:completeness}), so the PRISM selection preferentially misses the blue population rather than the LRDs. Second, sources that pass the optical compactness criterion but fail the UV compactness cut could in principle host a genuine hot ionizing source whose UV emission is outshone or blended with extended host-galaxy star formation at the spatial resolution and depth of the available imaging. Disentangling the central ionizing source from the host UV emission in these cases would require higher spatial resolution observations capable of resolving the UV morphology on sub-kpc scales. Our selection therefore provides a clean, conservative sample in which the UV properties can be unambiguously attributed to the central source, at the cost of potentially missing sources where the host contribution to the UV is non-negligible.

\begin{figure*}
\centering
\includegraphics[width=0.82\textwidth]{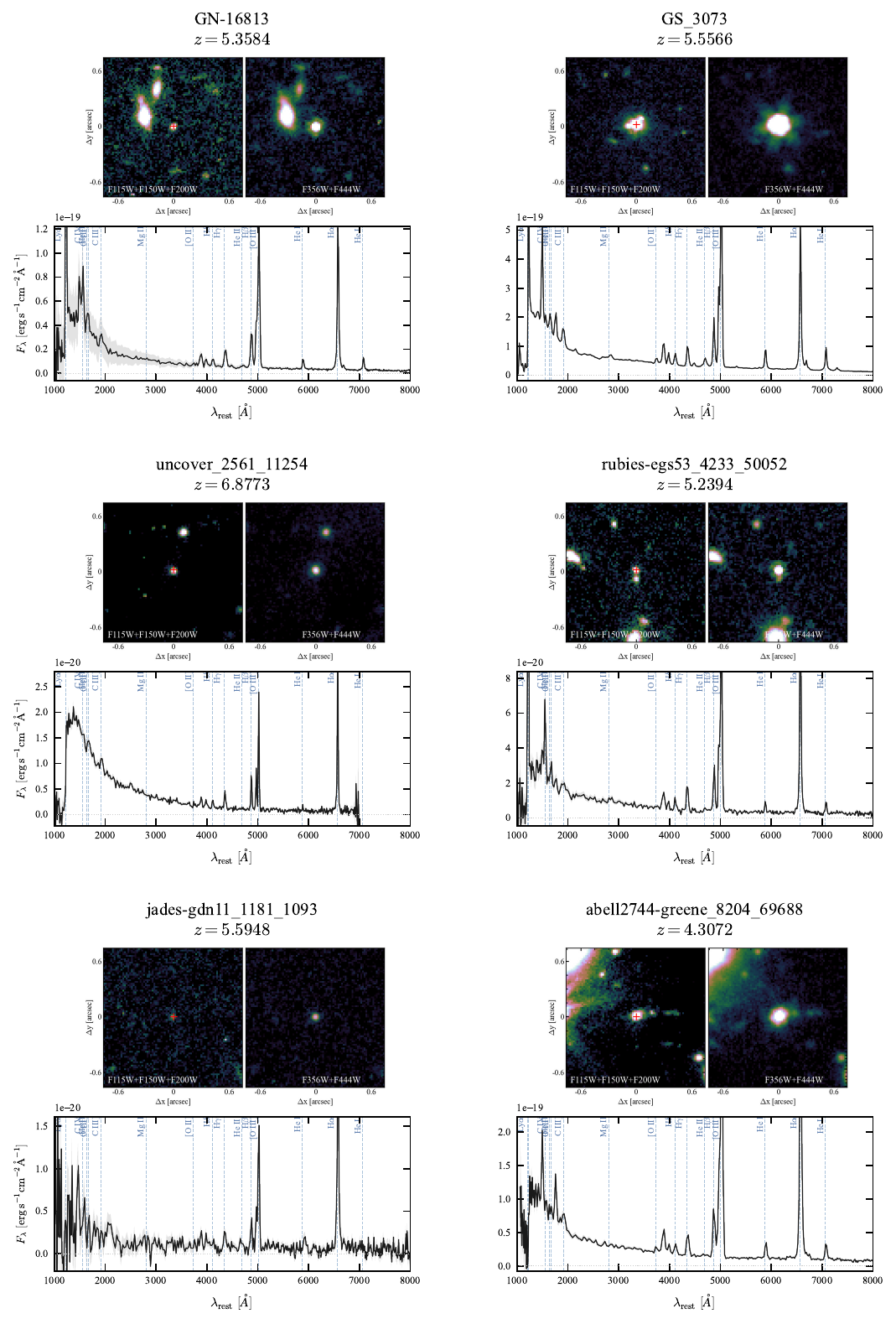}
\caption{NIRCam stamps and NIRSpec/PRISM spectra for 6 compact blue BLEs. For each source, the upper panel shows NIRCam composites in the rest-frame UV (F115W$+$F150W$+$F200W, left) and optical (F356W$+$F444W, right), with the optical centroid marked by a red cross on the UV stamp. The lower panel shows the PRISM spectrum (black) with the $1\sigma$ uncertainty band (grey). Vertical dashed lines mark the positions of prominent emission lines. Sources are ordered by increasing $\beta_{\rm UV}$ (blue to red from top to bottom), with the redshift indicated above each stamp. The remaining 14 sources are shown in Fig.~\ref{fig:stamps_spectra_appendix}.}
\label{fig:stamps_spectra}
\end{figure*}

We caution that any statement about LRDs dominating the broad-line population is luminosity-dependent. The PRISM broad-line selection is complete only for the most luminous emitters and, at fainter luminosities, more easily recovers the reddest sources (Sect.~\ref{sec:ha_selection}); the LRD fraction it returns is therefore representative only at the bright end and increasingly biased toward fainter luminosities. A robust characterization of the composition of the broad-line population below our completeness limit, will require a systematic analysis of the higher-resolution grating spectra, which can recover the fainter broad lines that the PRISM misses. The NIRCam images and the NIRSpec PRISM spectra for the 20 sources in the compact, blue BLE sample are shown in Fig.~\ref{fig:stamps_spectra} and Fig.~\ref{fig:stamps_spectra_appendix}.

\begin{table*}
\caption{Properties of the final sample of 20 compact, blue broad-line emitters.}
\label{tab:sample}
\centering
\small
\setlength{\tabcolsep}{4pt}
\begin{tabular}{lcllcccccc}
\hline\hline
Source & $z$ & R.A. [deg] & Dec. [deg] & $\beta_\mathrm{UV}$ & $\beta_\mathrm{opt}$ & $F_{\nu,\mathrm{red}}/F_{\nu,\mathrm{blue}}$ & $F(\mathrm{H}\alpha)/F(\mathrm{H}\beta)$ & $\mathcal{C}_\mathrm{opt}$ & $\mathcal{C}_\mathrm{UV}$ \\
\hline
jades-gdn09\_1181\_73488     & $4.133$ & $189.1974$ & $\phantom{-}62.1772$ & $-1.73 \pm 0.06$ & $\phantom{-}0.41 \pm 0.15$ & $2.60$ & $9.71$ & $1.63$ & $1.38$ \\
abell2744-greene\_8204\_69688  & $4.307$ & $3.5649$  & $-30.3482$      & $-2.12 \pm 0.04$ & $\phantom{-}0.39 \pm 0.17$ & $1.26$ & $6.29$ & $1.70$ & $1.37$ \\
capers-egs47\_6368\_19300     & $4.543$ & $215.0221$ & $\phantom{-}52.9208$ & $-1.53 \pm 0.08$ & $\phantom{-}0.14 \pm 0.17$ & $3.37$ & $14.31$ & $1.61$ & $0.92$ \\
nexus-obs3\_5105\_10835      & $4.650$ & $268.4377$ & $\phantom{-}65.1675$ & $-1.63 \pm 0.31$ & $-0.26 \pm 0.46$      & $4.67$ & $13.66$ & $1.61$ & $1.24$ \\
rubies-egs62\_4233\_42232    & $4.952$ & $214.8868$ & $\phantom{-}52.8554$ & $-1.50 \pm 0.28$ & $-0.64 \pm 0.37$      & $6.57$ & $11.45$ & $1.61$ & $1.60$ \\
uncover\_2561\_38108       & $4.969$ & $3.5300$  & $-30.3580$      & $-1.90 \pm 0.08$ & $\phantom{-}0.18 \pm 0.18$ & $2.54$ & $11.37$ & $1.54$ & $1.17$ \\
ceers-ddt\_2750\_1768       & $5.088$ & $214.9258$ & $\phantom{-}52.9457$ & $-2.11 \pm 0.25$ & $\phantom{-}0.46 \pm 0.36$ & $4.53$ & $11.84$ & $1.60$ & $0.90$ \\
rubies-egs53\_4233\_50052    & $5.239$ & $214.8235$ & $\phantom{-}52.8303$ & $-2.16 \pm 0.06$ & $-0.45 \pm 0.33$      & $1.29$ & $4.64$ & $1.63$ & $1.38$ \\
capers-egs61\_6368\_20952     & $5.283$ & $214.8802$ & $\phantom{-}52.8126$ & $-1.72 \pm 0.09$ & $\phantom{-}0.22 \pm 0.29$ & $2.20$ & $9.83$ & $1.63$ & $1.53$ \\
GN-16813             & $5.358$ & $189.1793$ & $\phantom{-}62.2925$ & $-2.48 \pm 0.20$ & $-0.83 \pm 0.42$      & $1.41$ & $3.46$ & $1.58$ & $1.12$ \\
GS\_3073             & $5.557$ & $53.0789$ & $-27.8842$      & $-2.25 \pm 0.02$ & $-0.20 \pm 0.01$      & $1.35$ & $3.21$ & $1.70$ & $1.37$ \\
rubies-uds23\_4233\_172350    & $5.581$ & $34.3690$ & $-5.1039$       & $-1.60 \pm 0.22$ & $-0.10 \pm 0.30$      & $4.35$ & $11.20$ & $1.57$ & $1.06$ \\
jades-gdn11\_1181\_1093     & $5.595$ & $189.1797$ & $\phantom{-}62.2246$ & $-2.14 \pm 0.25$ & $-0.57 \pm 0.92$      & $2.14$ & $7.49$ & $1.58$ & $1.19$ \\
macs1423\_1208\_4103248      & $5.775$ & $215.9612$ & $\phantom{-}24.0580$ & $-1.55 \pm 0.20$ & $\phantom{-}0.32 \pm 0.36$ & $6.32$ & $11.70$ & $1.58$ & $1.55$ \\
jades-gdn198\_1181\_38147    & $5.870$ & $189.2707$ & $\phantom{-}62.1484$ & $-1.53 \pm 0.10$ & $\phantom{-}0.35 \pm 0.39$ & $1.64$ & $7.53$ & $1.68$ & $1.33$ \\
gto-wide-uds12\_1215\_1259    & $5.962$ & $34.4223$ & $-5.2507$       & $-1.91 \pm 0.12$ & $-0.84 \pm 0.55$      & $2.03$ & $6.15$ & $1.60$ & $1.13$ \\
rubies-uds32\_4233\_966096    & $6.166$ & $34.4667$ & $-5.1009$       & $-2.04 \pm 0.23$ & $\phantom{-}0.05 \pm 1.40$ & $1.22$ & $18.77$ & $1.59$ & $1.19$ \\
capers-cos07\_6368\_35805     & $6.528$ & $150.0554$ & $\phantom{-}2.2916$  & $-2.06 \pm 0.03$ & $\phantom{-}0.28 \pm 0.17$ & $1.99$ & $7.65$ & $1.70$ & $1.56$ \\
jades-gdn2\_1181\_38509      & $6.672$ & $189.0915$ & $\phantom{-}62.2281$ & $-1.79 \pm 0.14$ & $-0.30 \pm 0.70$      & $1.64$ & $5.81$ & $1.70$ & $1.21$ \\
uncover\_2561\_11254       & $6.877$ & $3.5804$  & $-30.4050$      & $-2.24 \pm 0.03$ & $-0.62 \pm 0.44$      & $0.94$ & $4.25$ & $1.70$ & $1.14$ \\
\hline\hline
\end{tabular}
\tablefoot{
$F_{\nu,\mathrm{red}}/F_{\nu,\mathrm{blue}}$, the Balmer break strength, is defined as the ratio of the median flux density at $5400$--$5700$\,\AA\ to that at $3500$--$3650$\,\AA. $\mathcal{C} \equiv F_{0.2\arcsec}/F_{0.1\arcsec}$ is the compactness ratio measured in the rest-frame optical (F444W) and UV (F115W, F150W, or F200W; see Sect.~\ref{sec:compact_selection}). The Balmer decrement $F(\mathrm{H}\alpha)/F(\mathrm{H}\beta)$ is measured from the total line fluxes as described in Sect.~\ref{subsec:ha}.}
\end{table*}

\subsection{Example sources}

To illustrate the diversity of the sample, we briefly describe a subset of sources that are individually visible in Fig.~\ref{fig:stamps_spectra} and Fig.~\ref{fig:stamps_spectra_appendix}, and that connect to results in the literature.

\textbf{GN-16813} was first identified as a broad \ha\ emitter by \citet{Matthee2024}, who noted its compact morphology and broad Balmer emission. Subsequent IFU follow-up revealed a blue UV continuum with strong UV lines, unveiling its nature as a compact blue broad-line emitter rather than a classical LRD with a strong Balmer break. In the present sample, GN-16813 is the only source with medium-resolution grating spectroscopy covering the full rest-frame UV (from the DIVER program, GO-8018, PI Lin; see Sect.~\ref{subsec:uv}), making it, in terms of the UV, the best-characterized object in the sample.

\textbf{GS\_3073} was first studied in detail by \citet{Vanzella2010}, who detected strong \lya\ and \ion{N}{iv]}~$\lambda1486$ emission in the rest-frame UV. It was subsequently identified as an AGN at $z \approx 5.55$ by \citet{Grazian2020}, and by \citet{Ubler2023} via JWST/NIRSpec IFU spectroscopy, based on broad hydrogen and helium emission and a large \ion{He}{ii}~$\lambda4686$ equivalent width ($\mathrm{EW} \sim 20$\,\AA). \citet{Brazzini2026} identified it as the prototypical ``Rosetta Stone'' of the Little Blue Dot (LBD) class -- the blue, compact counterpart to LRDs -- finding no Balmer absorption, no time variability, and strong broad \ion{He}{ii} emission supporting their interpretation as BLR stratification and a relatively unobscured sight line to the central engine. In the present sample, its blue UV slope, negligible Balmer break, and low Balmer decrement ($F(H\alpha)/F(H\beta) = 3.21 \pm 0.05$) make it one of the least obscured sources and a natural anchor point for the population-level analysis in Sect.~\ref{sec:spectra}. We note that while GS\_3073 shows a very clumpy morphology in the rest-frame UV NIRCam image, the dominant UV emission is co-spatial with the H$\alpha$ emission. 

\textbf{rubies-egs53\_4233\_50052} corresponds to CEERS-2782, analyzed by \citet{Kocevski2023} as one of the first spectroscopically confirmed broad-line AGN candidates at $z > 5$ in the CEERS field. With $\beta_{\rm UV} = -2.16 \pm 0.06$ and detections of both \ion{N}{iv]} and \ion{C}{iv} in the PRISM spectrum (Sect.~\ref{subsec:uv}), it is among the most UV-bright sources in the sample. Similar to many LRDs \citep{Baggen2026}, it has a compact, nearby blue clump.

\textbf{abell2744-greene\_8204\_69688} is located behind the Abell~2744 cluster and is gravitationally lensed with a magnification of $\mu \sim 2$ \citep{Furtak2023a}. It was selected as a broad line emitters from ALT \citep{naidu2024} and presented in \cite{Matthee2025} and observed with JWST/NIRSpec MSA as part of GO program 8204 (PIs Greene \& Labb\'e), and is analyzed in \citet{Matthee2026}. Its UV morphology appears clumpy, suggestive of ongoing interactions. Its PRISM spectrum reveals exceptionally strong \ion{N}{iv]} and \ion{N}{iii]} emission.

Finally, we note that the selection boundary between compact blue BLEs and classical LRDs is not a sharp one, and a handful of sources in the sample occupy the transition region between the two populations. Sources such as \textbf{nexus-obs3\_5105\_10835}, \textbf{rubies-uds23\_4233\_172350}, and \textbf{macs1423\_1208\_4103248} display spectra that, while satisfying our selection criteria, show properties more reminiscent of LRDs: a redder UV continuum slope, a more pronounced Balmer break, and a higher Balmer decrement. As an illustrative example, \textbf{rubies-uds23\_4233\_172350} has been selected as LRD from \cite{Hviding2025} and shows $\beta_{\rm UV} \sim -1.60$ and $F(\mathrm{H}\alpha)/F(\mathrm{H}\beta) \sim$ 11.2, placing it close to the LRD locus in the $\beta_\mathrm{UV}$ versus $\beta_\mathrm{opt}$ diagram (Fig.~\ref{fig:beta_diagram}). Rather than excluding such borderline cases, we retain them in the sample as they naturally probe the continuum of properties bridging the two samples. This edge population is qualitatively consistent with the viewing-angle interpretation of Sect.~\ref{subsec:sirocco_results}, in which sources observed at larger $\theta$, intercepting more of the equatorial envelope, would naturally display redder slopes and stronger Balmer absorption, transitioning smoothly into the LRD regime, or is challenged by luminous host galaxy emission. In this case, the separation between LRDs and compact blue BLEs is placed at a relatively arbitrary threshold that reflects a specific gas column density \citep[e.g.,][]{Matthee2026}. 
Table~\ref{tab:appendix} lists all 20 sources, their observing programs, and the available data products.

\section{Spectral analysis}\label{sec:spectra}

In this section, we characterize the spectral properties of the 20 compact blue broad-line emitters selected in Sect.~\ref{sec:sample}, with the aim of constraining two quantities central to their role in cosmic reionization: the production rate of ionizing photons and the fraction that escape into the IGM. 

Constraining these quantities in this population is non-trivial. Standard calibrations of the ionizing photon production efficiency $\xi_{\rm ion}$ rely on \ha\ under the case B recombination assumption , but the compact blue BLEs may not satisfy this condition because high gas densities give rise to non case B conditions, for example due to a high Balmer optical depth \citep[e.g.,][]{Scarlata24}. Similarly, standard \lya-based estimates of the escape fraction and \ion{H}{i} column density rely on the line velocity offset and profile shape, which are inaccessible at the resolution of NIRSpec/PRISM (Sect.~\ref{subsec:lya}). We therefore use these diagnostics not to derive $\xi_{\rm ion}$ or $f_{\rm esc}^{\rm LyC}$ directly, but to build a consistent observational picture that motivates and constrains the self-consistent photoionization modeling of Sect.~\ref{sec:sirocco}.

\subsection{\texorpdfstring{\ha\ luminosity and Balmer decrement}{Ha luminosity and Balmer decrement}}
\label{subsec:ha}

The \ha\ luminosity and Balmer decrement $F(H\alpha)/F(H\beta)$ provide complementary constraints on the nebular conditions in the compact blue BLEs. While \ha\ traces the total recombination rate and sets a lower bound on the ionizing photon budget, the Balmer decrement encodes information on the physical conditions of the gas and the dust attenuation. In LRDs, Balmer decrements are typically very high \citep[with sample averages of $F(H\alpha)/F(H\beta) \approx 8.7$;][]{deGraaff2025b} and correlate with the Balmer break strength \citep{Matthee2026}, suggesting that both are driven by the same dense gaseous envelope through scattering and collisional (de-)excitation \citep{Torralba2026, deGraaff2025b,Chang2025}.
The blue UV slopes ($\beta_\mathrm{UV} \leq -1.5$) in our sample already suggest that the line of sight to the central source is relatively unobscured, though this alone cannot distinguish between dust attenuation and the dense gas scattering effects that characterize LRDs. Measuring the Balmer decrement provides a more direct handle on the dust and gas conditions along the line of sight, while acknowledging that elevated decrements can reflect both dust reddening and collisional effects in a dense medium, a degeneracy that the Balmer break strength and photoionization modeling of Sect.~\ref{sec:sirocco} help to break. We therefore measure both quantities from the PRISM spectra as follows.

The \ha\ emission line of each source was decomposed into a narrow and a broad Gaussian component following the criteria described in Sect.~\ref{sec:ha_selection}. In addition, we fit the \hb\ emission line with a single Gaussian plus a power-law continuum, using the same Levenberg--Marquardt minimization and LSF convolution described in Sect.~\ref{sec:ha_selection}. A single-component model is adopted for \hb\ because the broad component is not independently detected at sufficient S/N in the PRISM spectra; a two-component fit would introduce poorly constrained degeneracies between narrow and broad amplitudes at this spectral resolution. 
Following \citet{deGraaff2025b}, the Balmer decrement $F(H\alpha)/F(H\beta)$ is computed using the total \ha\ and \hb\ fluxes, as the spectral resolution of NIRSpec/PRISM does not allow a reliable decomposition into broad and narrow components.

\begin{figure}
\centering
\includegraphics[width=\columnwidth]{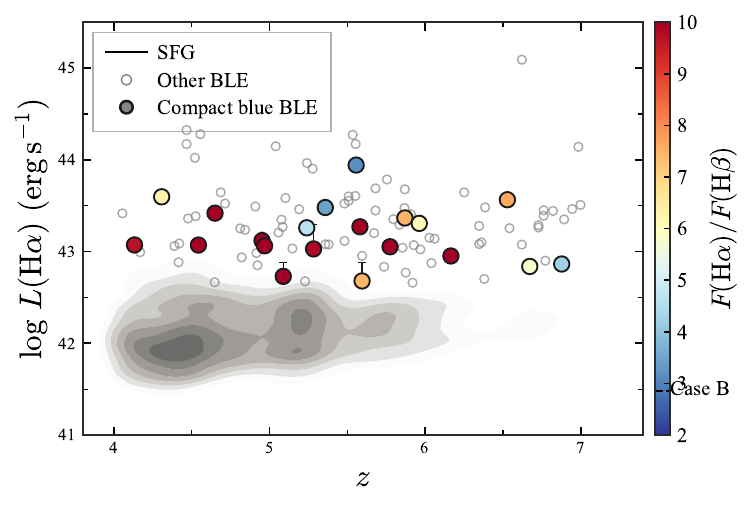}
\caption{\ha\ luminosity as a function of redshift. Filled circles show the 20 compact BLEs, color-coded by the observed Balmer decrement $F_{H\alpha}/F_{H\beta}$; the color scale goes from the Case B value of 2.86 (blue) to 10 (red). Open circles show the other BLE sample; gray contours the SFG sample from the \DJA\ compilation.}
\label{fig:ha_lum}
\end{figure}

Fig.~\ref{fig:ha_lum} shows the total \ha\ luminosity $L_{H\alpha}$ as a function of redshift for our 20 compact, blue BLEs (filled circles, color-coded by the observed Balmer decrement), compared with the other BLE sample (open circles) and the SFG sample from the \DJA\ compilation (gray contours). The compact blue BLEs span $\log(L_{\mathrm{H}\alpha}/\mathrm{erg\,s^{-1}}) \approx 42.7$--$43.9$, with a median of $43.2$, consistent with the broader BLE distribution (median $\log L_{\mathrm{H}\alpha} = 43.3$). Both samples lie systematically above the typical SFG luminosity at comparable redshifts  \citep[e.g.,][]{CoveloPaz24}.

Strikingly, every source in the sample has a Balmer decrement above the case B recombination value of $2.86$, spanning $F(H\alpha)/F(H\beta) \approx 3.2$--$18.8$ with a median of $8.7$, far above the value of 3.06 measured for low-redshift blue AGN by \citet{Dong2008}. The two sources with decrements closest to case B -- GS\_3073 ($F(H\alpha)/F(H\beta) \sim 3.21$) and GN-16813 ($\sim 3.46$) -- are consistent with a relatively clear nebular sight line, in agreement with the strong \lya\ detection and low \ion{H}{i} column inferred for GN-16813 (Sect.~\ref{subsec:lya} and ~\ref{subsec:uv}). The majority of the sample shows substantially elevated values: 16 out of 20 sources have $F(H\alpha)/F(H\beta) > 5$, and 8 exceed $F(H\alpha)/F(H\beta) > 10$. These high decrements point to a dense, partially covering gas envelope that preferentially drives the Balmer decrements away from case B, consistent with the geometry proposed for LRDs \citep[e.g.,][]{Sneppen2026a, Matthee2026}. This is discussed in the context of our photo-ionization modeling in Sect.~\ref{subsec:stack}.
 
\subsection{Optical emission line diagnostics}
\label{subsec:optical}

Classical BPT diagrams \citep{Baldwin1981} lose discriminating power at high redshift: lower metallicities and harder ionizing fields push both star-forming galaxies and AGN narrow-line regions toward overlapping loci, making AGN selection based on rest-frame optical diagnostics ambiguous \citep[e.g.,][]{Harikane2023, Ubler2023, Scholtz2025}. We instead exploit the medium- and high-resolution grating spectra, which resolve emission lines blended at PRISM resolution, searching for [\ion{O}{ii}]~$\lambda \lambda 3727,3729$, [\ion{O}{iii}]~$\lambda\lambda 4959,5007$, H$\gamma$, and the auroral line [\ion{O}{iii}]~$\lambda 4363$. Of particular interest is the latter, recently proposed as an AGN diagnostic at high redshift: \citet{Mazzolari2024} showed that [\ion{O}{iii}]~$\lambda 4363$/H$\gamma$ is significantly elevated in AGN due to the harder ionizing spectrum driving the gas to higher temperatures. At PRISM resolution [\ion{O}{ii}]~$\lambda 3727$ and H$\gamma$ are blended with neighboring lines, restricting this analysis to the 4/20 sources with complete grating coverage (see Table \ref{tab:appendix}). Line fluxes and equivalent widths are measured with \texttt{unite} \citep{unite_code}.

\begin{figure}
\centering
\includegraphics[width=\columnwidth]{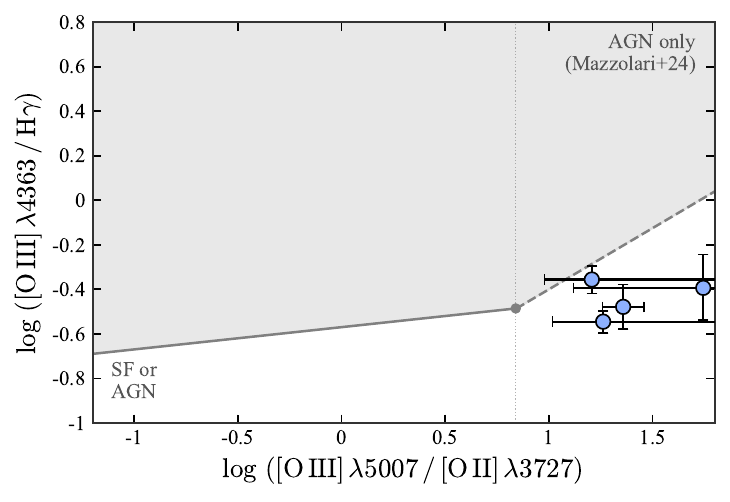}
\caption{Optical AGN diagnostic diagram following \citet{Mazzolari2024}. $y$-axis: auroral-to-Balmer ratio $\log([\ion{O}{iii}]\,\lambda 4363/\mathrm{H}\gamma)$; $x$-axis: ionization-sensitive ratio $\log([\ion{O}{iii}]\,\lambda 5007/[\ion{O}{ii}]\,\lambda 3727)$. Filled circles show the compact BLEs with grating coverage for all the four lines.}
\label{fig:optical_bpt}
\end{figure}

Fig.~\ref{fig:optical_bpt} shows the [\ion{O}{iii}]~$\lambda 4363$/H$\gamma$ versus [\ion{O}{iii}]~$\lambda 5007$/ [\ion{O}{ii}]~$\lambda 3727$ diagnostic diagram \citep{Mazzolari2024} for compact BLEs with grating coverage and significant ($\mathrm{S/N}\geq3$) detections in all four lines. The demarcation line from \citet{Mazzolari2024}, based on the photoionization models of \citet{Gutkin2016} and \citet{Feltre2016}, separates the AGN-only region (upper left) from the region accessible to both star-forming galaxies and AGN. Our sources predominantly occupy the latter region, consistent with star-forming galaxies or composite systems rather than AGN.

The fact that our sources are not identified as AGN by this optical diagnostic does not contradict their broad-line nature. Rather, it suggests that the [\ion{O}{iii}]~$\lambda 4363$ emission in these objects is not strongly enhanced relative to H$\gamma$, which may reflect low gas-phase metallicity (although this is probably only relevant at metallicities below $\approx5$\% solar that are rare at these redshifts, e.g. \citealt{Kotiwale26}) or a softer ionizing spectra than typical AGN models, yielding more modest temperature increases, or some contribution from star-formation-powered photoionization to the narrow-line emission. Regardless, none of the rest-frame optical emission-line diagnostics convincingly shows that the broad-line sources are powered by a classical AGN spectrum.

\subsection{\texorpdfstring{\lya\ detection}{lya detection}}
\label{subsec:lya}

\lya\ is a powerful probe of the neutral gas covering fraction in the ISM. Its emerging equivalent width, escape fraction and velocity offset are sensitive to the \ion{H}{i} column density and geometry along the line of sight \citep[e.g.,][]{Tang26} and therefore also to the ionizing photon escape fraction \citep[e.g.,][]{Verhamme2017}. At the resolution of NIRSpec/PRISM, however, the \lya\ line profile is severely blurred by the instrumental LSF ($\mathrm{FWHM} \sim 10^4$\,km\,s$^{-1}$ at $z \sim 4$--$7$, where $R \sim 30$--$50$), making it impossible to characterize the line profile, measure velocity offsets, or detect lines with $\mathrm{EW}_0 \lesssim 20$\,\AA\ \citep{Jones2024}.
PRISM-based \lya\ measurements are therefore restricted to integrated line flux and equivalent width: they are sufficient to identify \lya\ emitters and quantify the LAE fraction across the sample, but insufficient to constrain the escape kinematics or the \ion{H}{i} column in individual sources. With these caveats in mind, we fit the \lya\ emission line of all analyzed sources (both SFGs and BLEs) following the method of \citet{Napolitano2024}.

\begin{table*}
\caption{\lya\ measurements for the 20 compact blue broad-line emitters, sorted by $M_{\rm UV}$.}
\label{tab:lya}
\centering
\small
\setlength{\tabcolsep}{4pt}
\begin{tabular}{lcccc}
\hline\hline
Source &
$M_{\rm UV}$ &
$F_\mathrm{Ly\alpha}$ &
$\mathrm{EW}_0$ &
S/N \\
 &
$[\mathrm{AB \ mag}]$ &
$[10^{-18}\,\mathrm{erg\,s^{-1}cm^{-2}}]$ &
$[\mathrm{\AA}]$ & \\
\hline
GS\_3073 & $-24.54 \pm 0.01$ & $105.84 \pm 0.27$ & $ 65.4 \pm 0.2$ & 391.5 \\
capers-cos07-v4\_6368\_35805  & $-22.86 \pm 0.03$ & $ 8.98 \pm 0.51$ & $ 33.1 \pm 1.9$ & 17.7 \\
GN-16813     & $-22.83 \pm 0.16$ & $ 57.37 \pm 4.87$ & $151.6 \pm 12.9$ & 11.8 \\
uncover-v4\_2561\_11254   & $-22.50 \pm 0.03$ & $< 0.48$   & $< 1.4$   & 1.2 \\
gto-wide-uds12-v4\_1215\_1259  & $-22.41 \pm 0.09$ & $ 32.63 \pm 1.48$ & $161.0 \pm 7.3$ & 22.1 \\
rubies-egs53-v4\_4233\_50052  & $-22.25 \pm 0.04$ & $ 51.03 \pm 1.14$ & $235.1 \pm 5.2$ & 44.9 \\
abell2744-greene-v4\_8204\_69688 & $-21.82 \pm 0.03$ & $< 1.49$   & $< 5.8$   & 1.0 \\
rubies-uds32-v4\_4233\_966096  & $-21.70 \pm 0.16$ & $< 1.28$   & $< 15.7$   & 1.0 \\
jades-gdn198-v4\_1181\_38147  & $-21.69 \pm 0.10$ & $ 2.74 \pm 0.74$ & $ 23.8 \pm 6.5$ & 3.7 \\
jades-gdn2-v4\_1181\_38509  & $-21.30 \pm 0.11$ & $ 1.51 \pm 0.44$ & $ 28.9 \pm 8.3$ & 3.5 \\
capers-egs61-v4\_6368\_20952  & $-21.05 \pm 0.11$ & $ 4.81 \pm 0.65$ & $ 41.5 \pm 5.6$ & 7.4 \\
uncover-v4\_2561\_38108   & $-20.81 \pm 0.08$ & $ 2.46 \pm 0.47$ & $ 27.0 \pm 5.2$ & 5.2 \\
rubies-uds23-v4\_4233\_172350  & $-20.78 \pm 0.27$ & $ 6.05 \pm 1.00$ & $111.8 \pm 18.5$ & 6.0 \\
macs1423-v4\_1208\_4103248  & $-20.77 \pm 0.24$ & $< 2.62$   & $< 18.3$   & 2.9 \\
capers-egs47-v4\_6368\_19300  & $-20.48 \pm 0.07$ & $< 1.60$   & $< 12.9$   & 2.8 \\
jades-gdn09-v4\_1181\_73488  & $-20.43 \pm 0.07$ & $ 30.95 \pm 1.21$ & $225.3 \pm 8.8$ & 25.5 \\
rubies-egs62-v4\_4233\_42232  & $-20.38 \pm 0.34$ & $< 2.13$   & \ldots   & 1.1 \\
nexus-obs3-v4\_5105\_10835  & $-20.24 \pm 0.41$ & $ 10.48 \pm 2.80$ & \ldots   & 3.7 \\
jades-gdn11-v4\_1181\_1093  & $-20.23 \pm 0.25$ & $< 1.05$   & $< 20.7$   & 1.2 \\
ceers-ddt-v4\_2750\_1768   & $-19.34 \pm 0.26$ & $< 0.71$   & $< 23.9$   & 1.2 \\
\hline
\end{tabular}
\tablefoot{Upper limits ($3\sigma$) are given for sources with $\mathrm{S/N} < 3$. For nexus-obs3 v4\_5105\_10835 and rubies-egs62-v4\_4233\_42232, the continuum redward of Ly$\alpha$ is unresolved in the PRISM spectrum, preventing a reliable EW measurement (\ldots).}
\end{table*}

The line profile is modeled as a Gaussian convolved with the NIRSpec/PRISM instrumental line spread function, superimposed on a step-function continuum that accounts for the sharp break in \lya\ opacity blueward of the line. The blue-side continuum level is fixed to zero, consistent with full IGM absorption at $z \gtrsim 4$, while the red-side continuum is modeled as a linear function fitted over the rest-frame window $1260$--$1900$\,\AA. The continuum level at the \lya\ wavelength, $C_\mathrm{Ly\alpha}$, is evaluated by averaging the best-fit linear model over $1216$--$1240$\,\AA. The LSF is approximated as a Gaussian with $\sigma_\mathrm{LSF}$ derived from the point-source resolution curve evaluated at the observed \lya\ wavelength. The free parameters (line amplitude and FWHM) are sampled with the affine-invariant MCMC ensemble sampler \textsc{emcee} \citep{ForemanMackey2013}, using $N_\mathrm{walkers} = 64$ walkers, $N_\mathrm{steps} = 2000$ steps and a burn-in of 500 steps. Parameter uncertainties correspond to the 16th--84th percentile range of the marginalized posteriors.

A \lya\ detection is claimed at $\mathrm{S/N} \geq 3$, with the peak position searched within a velocity window of $-300$ to $+1000$\,km\,s$^{-1}$ relative to the systemic \lya\ wavelength. The rest-frame equivalent width is computed as
$
 \mathrm{EW}_0(\mathrm{Ly}\alpha) =
 F_\mathrm{Ly\alpha}/[(1+z)\,C_\mathrm{Ly\alpha}], $
where $C_\mathrm{Ly\alpha}$ is the continuum flux density evaluated redward of the line from the best-fit linear model. For non-detections, a $3\sigma$ upper limit on the EW is derived from the local noise at the expected \lya\ peak position and the LSF width following
$
 \mathrm{EW}_{0,\,\mathrm{lim}} = \sqrt{2\pi}\,\sigma_\mathrm{LSF}\,\sigma_\mathrm{noise}/
   [(1+z)\,C_\mathrm{Ly\alpha}].$

To ensure the LAE fractions are not biased by sources whose continuum is too faint to constrain the EW, we apply an injection-recovery completeness test to all non-detections. We inject a synthetic \lya\ line with $\mathrm{EW}_0 = 25$\,\AA\ -- the LAE threshold -- into each spectrum at $+500$\,km\,s$^{-1}$ redward of the systemic wavelength, perturb the spectrum with $N_\mathrm{trials} = 100$ independent noise realizations, and attempt to recover the injected line with the full MCMC fitter. Non-detections for which the median recovered $\mathrm{S/N} \geq 3$ are classified as genuine non-LAEs and retained in the analysis; those for which the injection is not recovered are classified as incomplete and excluded from the denominator when computing $X_{\rm Ly\alpha}$.

\begin{figure}
\centering
\includegraphics[width=\columnwidth]{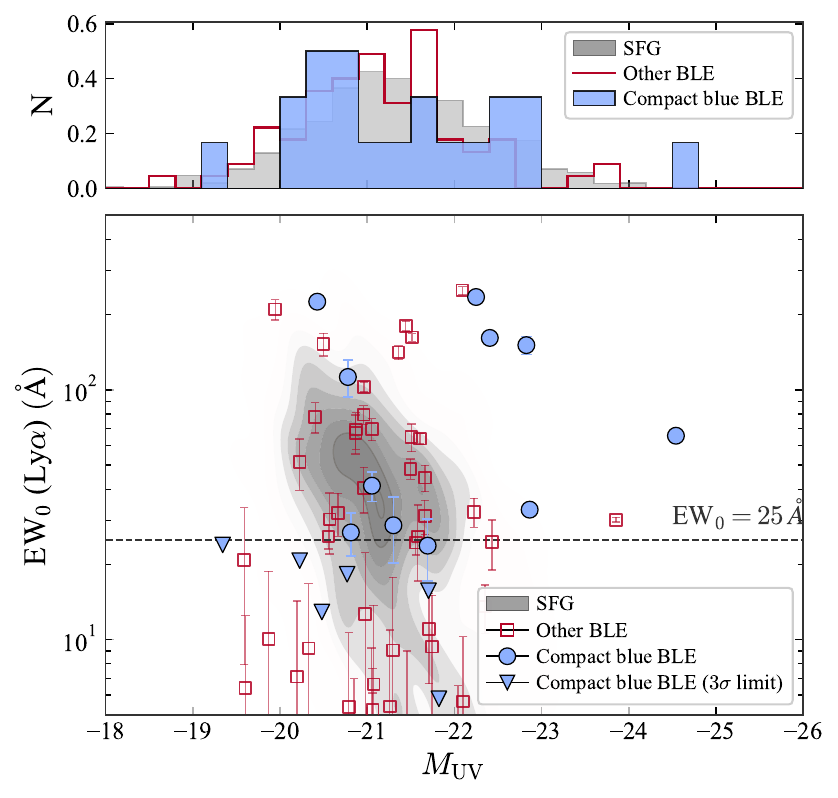}
\caption{\lya\ rest-frame equivalent width $\mathrm{EW}_0$ as a function of absolute UV magnitude $M_\mathrm{UV}$. Filled circles show compact blue BLEs with significant detection ($\mathrm{S/N} \geq 3$), with $1\sigma$ error bars; downward triangles are $3\sigma$ upper limits. Open squares show the other BLE sample with $1\sigma$ error bars. Grey filled contours show the SFG comparison sample. The $M_{\rm UV}$ distribution for the three populations is presented in the upper panel.}
\label{fig:lya_ew}
\end{figure}

Among the 20 compact blue BLEs, 12 are detected in \lya\ with $\mathrm{S/N} \geq 3$, of which 11 satisfy $\mathrm{EW}_0 > 25$\,\AA, yielding an overall \lya\ emitter (LAE) fraction of $55$\%. The detected equivalent widths span $\mathrm{EW}_0 \approx 24$--$235$\,\AA, with a median of $\approx 65$\,\AA.

Fig.~\ref{fig:lya_ew} shows $\mathrm{EW}_0$ as a function of $M_\mathrm{UV}$ for all three samples. At fixed $M_\mathrm{UV}$, the compact blue BLEs show systematically elevated $\mathrm{EW}_0$ relative to the SFG population across the full redshift range $4 \lesssim z \lesssim 7$ spanned by the sample (median $z \approx 5.5$), consistent with a stronger ionizing continuum boosting intrinsic \lya\ production and/or a higher \lya\ escape fraction.

\begin{figure}
\centering
\includegraphics[width=0.9\columnwidth]{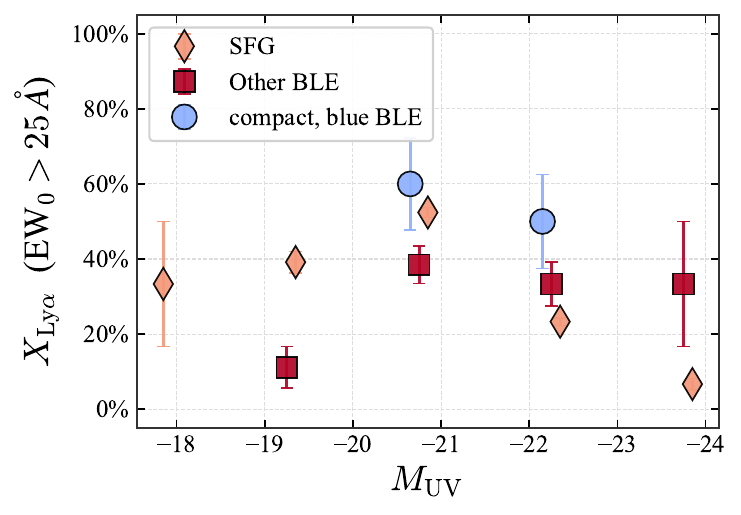}
\caption{\lya\ emitter fraction $X_{\rm Ly\alpha}$ (fraction of sources with $\mathrm{EW}_0 > 25$\,\AA) as a function of $M_{\rm UV}$, in bins of $\Delta M_{\rm UV} = 1.5$\,mag. Bins containing fewer than 3 sources are omitted. Error bars represent Poisson uncertainties for SFGs and small-number statistics for the other two samples. Orange diamonds: SFG sample. Dark red squares: other BLE sample (excluding compact blue BLEs). Blue circles: compact blue BLE subsample.}

\label{fig:lya_fraction}
\end{figure}

\begin{figure}
\centering
\includegraphics[width=0.9\columnwidth]{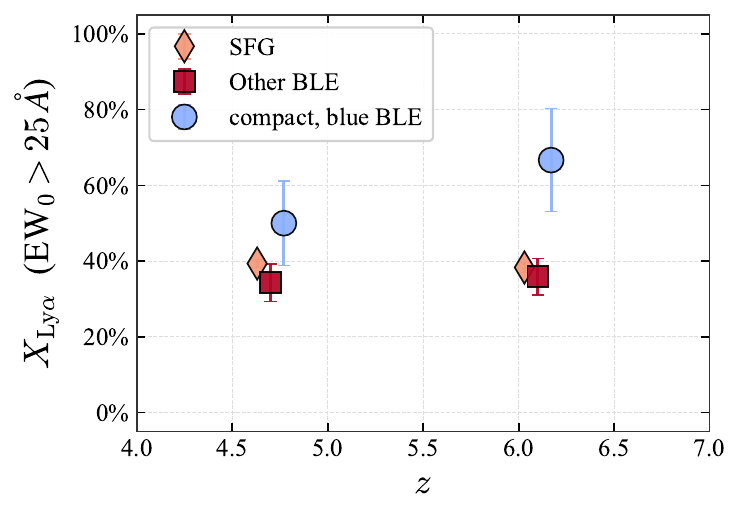}
\caption{\lya\ emitter fraction $X_{\rm Ly\alpha}$ (fraction of sources with $\mathrm{EW}_0 > 25$\,\AA) as a function of redshift, in bins of $\Delta z \sim 1.5$. Symbols are as in Fig.~\ref{fig:lya_fraction}.}
\label{fig:lya_fraction_z}
\end{figure}

Fig.~\ref{fig:lya_fraction} quantifies this enhancement via the \lya\ emitter fraction $X_{\rm Ly\alpha}$, i.e. the fraction of sources with $\mathrm{EW}_0 > 25$\,\AA, as a function of $M_{\rm UV}$ in bins of $\Delta M_{\rm UV} = 1.5$\,mag. Poisson errors are used for the SFG sample; small-number statistics \citep{Gehrels1986} for the BLEs; bins with fewer than 3 sources are omitted. The SFG sample spans $-18 \lesssim M_{\rm UV} \lesssim -24$, with $X_{\rm Ly\alpha}$ ranging from $\approx 7$\% at the bright end to a peak of $\approx 53$\% near $M_{\rm UV} \approx -21$, in broad agreement with published LAE fractions for UV-selected galaxies at these redshifts \citep[e.g.,][]{Stark2011, DeBarros2017,Napolitano2024}. The other BLE sample spans $-19 \lesssim M_{\rm UV} \lesssim -24$ and shows systematically elevated fractions of $\approx 11$--$39$\% across this range. The compact blue BLEs stand out most strikingly: both bins in which the subsample is represented reach $X_{\rm Ly\alpha} \gtrsim 50$\%, consistently above both the SFG and other BLE levels at matched luminosity. The redshift dependence of $X_{\rm Ly\alpha}$ is shown in Fig.~\ref{fig:lya_fraction_z}, where both the SFG and other BLE populations show little evolution between $z \approx 4.6$ and $z \approx 6.1$, with fractions of $\approx 35$--$39$\% in both redshift bins \citep[e.g.,][]{Stark2011, DeBarros2017, Jones2024}. The compact blue BLEs show higher fractions in both bins, rising from $\approx 50$\% at $z \approx 4.8$ to $\approx 67$\% at $z \approx 6.2$, though the large uncertainties preclude a firm conclusion about redshift evolution within this subsample.

\subsection{UV emission lines}
\label{subsec:uv}
Rest-frame UV emission lines provide a direct probe of the ionization conditions and hardness of the ionizing continuum, complementing the \lya\ and \ha\ diagnostics discussed above. We search for the brightest UV metal and helium lines in the NIRSpec/PRISM spectra of the 20 compact, blue BLEs: \ion{N}{iv]}\,$\lambda\lambda1483,1487$, \ion{C}{iv}\,$\lambda\lambda1548,1551$, \ion{He}{ii}\,$\lambda1640$, \ion{O}{iii]}\,$\lambda\lambda1661,1666$, and \ion{C}{iii]}\,$\lambda\lambda1907,1909$. Each line is fitted with \texttt{unite} \citep{unite_code}. A detection is claimed at $\mathrm{S/N} \geq 3$.

\ion{O}{iii]}\,$\lambda\lambda1661,1666$ is the most commonly detected UV line in our sample, recovered in 4 out of 20 sources; we note however that at the PRISM resolution this line is blended with \ion{He}{ii}\,$\lambda1640$. \ion{N}{iv]}\,$\lambda\lambda1483,1487$, which requires ionizing photons above $47.4$\,eV, is detected in 3 sources. \ion{C}{iv}\,$\lambda\lambda1548,1551$ and \ion{C}{iii]}\,$\lambda\lambda1907,1909$ are each detected in 2 sources. The overall detection rates are low, as expected given the PRISM resolution and the modest per-pixel S/N ratio of these faint sources; confirmed detections are found exclusively among the sources with the highest continuum S/N in the sample. However, we note that PRISM spectra of {\it typical} star-forming galaxies do not show such UV lines \citep[e.g.,][]{Hayes25}.

The UV line properties collectively reinforce the picture established by \lya\ and broad \ha. The simultaneous detection of \ion{N}{iv]} and \ion{C}{iv} in rubies-egs53-v4\_4233\_50052 -- both requiring $>47$\,eV photons -- and of \ion{N}{iv]} at large equivalent width ($\mathrm{EW}_0 > 20$\,\AA) in abell2744-greene-v4\_8204\_69688 places these sources in the AGN locus of the \ion{C}{iv} EW diagnostic \citep{Nakajima2018}. 
Notably, \ion{N}{iv]} detections at high equivalent width are rare in the general galaxy population and suggest nitrogen enhancement \citep[e.g.,][]{Cameron23, Zhu2026}, as also reported for GN-z11 \citep{Maiolino2024} and other extreme high-$z$ systems \citep{Berg2025}. However, these high-ionization features are absent in the stacked spectrum (Sect.~\ref{subsec:stack}), suggesting that the UV emission in the sample as a whole reflects a superposition of host galaxy and central engine contributions, with the AGN UV signature emerging only in the highest-S/N individual sources.

At the resolution of NIRSpec/PRISM ($R \sim 50$--$100$), UV doublets are unresolved, \ion{O}{iii]} is blended with \ion{He}{ii}, and lines with $\mathrm{EW}_0 \lesssim 20$\,\AA\ fall below the effective detection threshold even at moderate continuum S/N ratio \citep[e.g.,][]{Jones2024}. Medium- and high-resolution spectroscopy is therefore required to resolve blended features, detect weaker UV lines, identify ISM absorption, and disentangle AGN, nebular, and stellar-wind contributions \citep[e.g.,][]{Vasan26}.

\subsection{Overall spectral shape}
\label{subsec:stack}
To illustrate the population-averaged spectral properties of the compact, blue BLEs, we construct a median rest-frame stack of all 20 sources. For each source, the observed-frame PRISM spectrum is shifted to rest-frame wavelengths using the spectroscopic redshift and normalized over the rest-frame window $5200$--$5500$\,\AA. All spectra are then interpolated onto a common rest-frame wavelength grid with spacing $\Delta\lambda_{\rm rest} = \Delta\lambda_{\rm inst}/(1+z_{\rm median})$, where $\Delta\lambda_{\rm inst}$ is the instrumental pixel scale. The 
stacked spectrum is computed as a weighted median, with weights derived from the per-pixel S/N ratio estimated via a rolling median, in order to minimize the impact of outliers and spectral artifacts. A weighted mean stack and its standard error are computed in parallel for comparison, and the results are consistent.

\begin{figure*}
\centering
\includegraphics[width=0.95\textwidth]{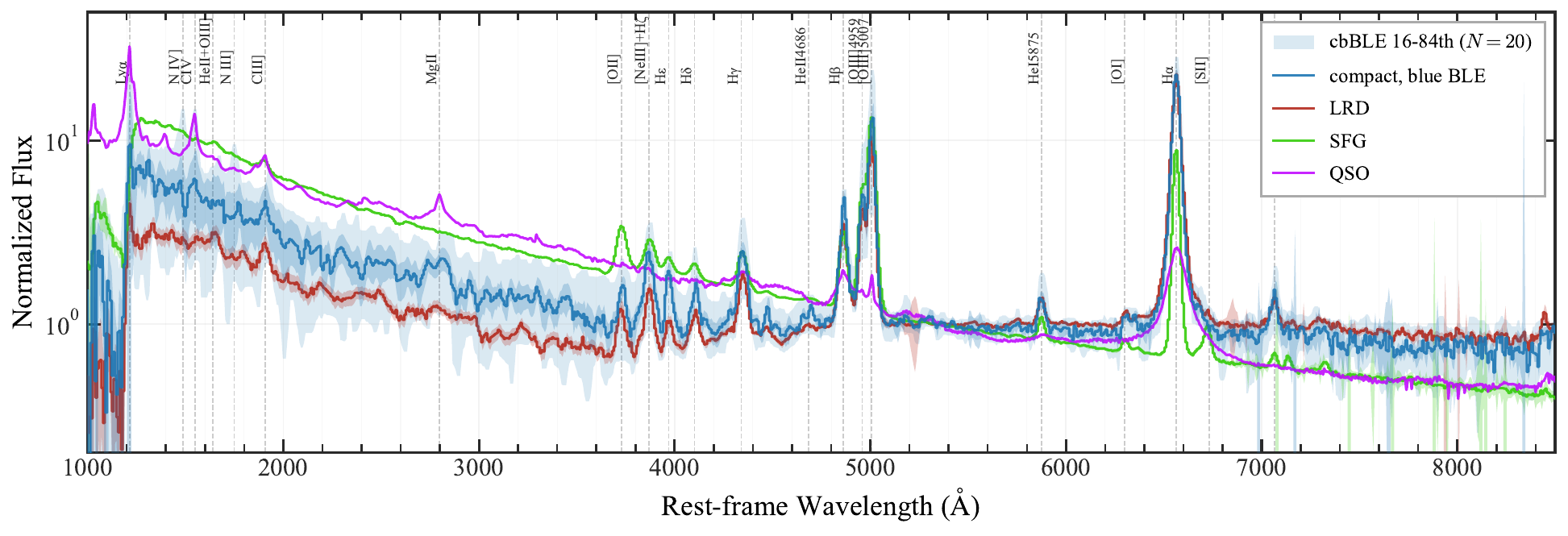}
\caption{Rest-frame stacked spectra of high-redshift galaxy populations. The compact, blue BLEs' median stack is shown in blue, the green line shows the median stack of star-forming galaxies at $z = 4-7$ from DJA. The shaded region indicating $\pm 1\sigma$. The red line shows the median stack of the LRDs from the ``Other BLE'' sample. The purple line shows the median stack of QSO at $z \sim 1.5$ from \cite{Selsing2016}. Vertical dashed gray lines mark prominent emission lines. All spectra are normalized between 5200--5500 \AA\ to facilitate comparison of spectral shapes. The compact, blue BLEs' stack exhibits distinct spectral features compared to the star-forming population, particularly in the strength and profile of emission lines such as \ha, \hb, and He~\textsc{i}, suggesting different physical conditions and/or dominant ionization mechanisms.}
\label{fig:stack}
\end{figure*}

The resulting stack is shown in Fig.~\ref{fig:stack}, together with equivalent stacks of the LRDs from the ``other BLE'' sample and the SFG comparison sample constructed with the same procedure, and a QSO template at $1 < z < 2.1 $ \citep{Selsing2016} for reference. The rest-frame UV continuum exhibits a consistently blue slope with $\beta_{\rm UV} = -1.72 \pm 0.04$ for the median stack. For comparison, the median stack of the LRDs has a $\beta_{\rm UV} = -1.46 \pm 0.04$ \citep[in agreement with][]{Ando2026}. 

Despite sharing several spectral features with broad-line quasars -- including broad Balmer emission and strong high-ionization lines -- compact blue BLEs are fundamentally distinguished by their significantly lower luminosities, and differences in the optical $\beta$ slopes. Moreover, we do not identify prominent {\it broad} lines in the UV such as \lya, \civ, \ciii\ or \ion{Mg}{ii}, which are hallmarks of classical broad-line quasar spectra. In the rest-frame optical stack, the compact blue BLEs exhibit $\mathrm{EW}_0(\mathrm{H}\alpha) \approx 900$\,\AA, significantly larger than the $\sim 200$\,\AA\ typical of luminous quasars \citep[e.g.,][]{VandenBerk2001}. Compared to the SFG population, both BLE samples are distinguished by the absence of [\ion{S}{ii}]\,$\lambda\lambda6718,6733$ emission ($3\sigma$ upper limits of $\mathrm{EW}_0 < 7.4$ and $4.6$\,\AA\ for the compact blue BLE and LRD stacks respectively, versus detections of $\mathrm{EW}_0 \approx 21$\,\AA\ in the SFG stack), consistent with a harder ionizing continuum suppressing low-ionization forbidden line emission. The compact blue BLE stack also shows strong \ion{He}{i}~$\lambda5875$ ($\mathrm{EW}_0 = 15.9 \pm 4.3$\,\AA) and \ion{He}{i}~$\lambda7065$ ($27.3 \pm 6.5$\,\AA) emission -- notably stronger than in the SFG stack ($12.5 \pm 0.7$ and $5.1 \pm 2.0$\,\AA\ respectively).

The rest-frame optical part of the compact blue BLE and LRD stacks are strikingly similar, sharing comparable \ha\ equivalent widths, Balmer line ratios, optical $\beta$ slopes, and \ion{He}{i}~$\lambda5875$ and \ion{He}{i}~$\lambda7065$ line strengths. This resemblance suggests that the compact blue BLEs and LRDs may share a common powering source, with the observed differences in UV properties reflecting variations in covering gas that may be due to differences in the viewing angle or in evolutionary phases of the dense envelope \citep{Matthee2026,Sneppen2026b}, rather than a fundamentally distinct physical nature.  We explore this connection in detail in Sect. $\ref{sec:sirocco}$.

\section{Photoionization modeling with \textsc{Sirocco}}
\label{sec:sirocco}

Considering the spectral similarities between compact blue BLEs and LRDs discussed in Sect.~\ref{subsec:stack}, together with their similar X-Ray faintness and lack of variability \citep{Brazzini2026}, compact blue BLEs have been proposed to represent the counterparts of LRDs viewed from sight lines in the envelope with lower column densities \citep{Matthee2026, Sneppen2026a}, where the UV continuum and associated high-ionization lines can escape. This is supported by their elevated LAE fractions relative to both the BLE and SFG baselines at matched luminosity (Fig.~\ref{fig:lya_fraction}), and their large \lya\ equivalent widths (Fig.~\ref{fig:lya_ew}), all pointing to low \ion{H}{i} columns along the line of sight. 

If compact blue BLEs and LRDs are primarily distinguished by envelope geometry and line-of-sight opacity, standard diagnostics of $\xi_{\rm ion}$ and $f_{\rm esc}^{\rm LyC}$ may not be directly applicable. They are calibrated on typical star-forming galaxies and rest on assumptions most critically case B recombination and an optically thin Lyman series -- that are unlikely to hold when a dense gas envelope surrounds the central ionizing source. In this geometry, ionizing photons are reprocessed before escaping, and the emergent line and continuum fluxes reflect the combined effects of column density, covering factor, and viewing angle rather than the intrinsic photon budget \citep{Chang2026}. 

For this reason, we use 3D\footnote{\textsc{Sirocco} is a 2.5-dimensional code, where the coordinate grid is 2D and symmetric about the z-axis. However, photon transport occurs in 3D, allowing rotational motion around the y-axis.} Monte Carlo radiative transfer simulations with \textsc{Sirocco} \citep{Matthews2025} to model the ionizing photon production and escape directly. While \textsc{Cloudy} remains the standard tool for photoionization modeling, it assumes a static medium and cannot self-consistently predict resolved emission-line profiles, since it lacks a treatment of the gas kinematics. \textsc{Sirocco} simultaneously reproduces the continuum and line-profile shapes for a given gas configuration, retaining the flexibility to explore different geometries, density structures, and clumping factors.

The constraints on the ionizing spectrum from the SED shape (Sect.~\ref{subsec:stack}) and on the gas geometry from the Balmer line profiles together provide sufficient information to set up the simulations (Sect.~\ref{subsec:sirocco_setup}). \textsc{Sirocco} employs a Monte Carlo radiative transfer framework specifically designed to treat optically thick, multi-dimensional flows. The code self-consistently solves the ionization structure, radiation field, electron column density, recombination line strengths and widths, extent of partially ionized material, and the strength of Balmer absorption lines and breaks throughout the gas cocoon \citep{Parkinson2025,Matthews2025}. However, \textsc{Sirocco} has a number of limitations \citep{Matthews2025}. Line transfer is treated in the Sobolev approximation which assumes that the wind has a sufficiently large velocity gradient that a photon traversing the wind sweeps through the resonance wavelength of any given transition on a timescale much shorter than the transition lifetime. As a result, only permitted transitions can be excited: the photon passes through resonance before a forbidden or semi-forbidden transition, which requires a comparatively long interaction time, can occur. This naturally suppresses forbidden line emission such as [\ion{O}{iii}]$\lambda\lambda$4959,5007, which are therefore not considered in the \textsc{Sirocco} wind model. The approximation neglects thermal and microturbulent broadening of bound-bound transitions, but may only introduce minor uncertainties in the cores of optically thick lines. Finally, the code assumes thermal and ionization equilibrium throughout the wind, which may not hold in the fastest-moving regions of the outflow where the local dynamical timescale is shorter than the recombination timescale.

\subsection{Model setup and physical configuration}
\label{subsec:sirocco_setup}

To identify a physical configuration capable of reproducing the observed spectral properties of the compact blue BLE sample, we explore a grid of \textsc{Sirocco} radiative transfer models varying the wind geometry, column density, and ionizing source properties (the full grid is described in Sect.~\ref{subsec:fesc_bb_calibration} and Appendix~\ref{appendix:sirocco_grid}). Here we present the single best-fit model from this exploration, selected by minimizing the reduced $\chi^2$ between the model and the stacked compact blue BLE spectrum, jointly across the four Balmer-break bins presented in \cite{Matthee2026} and all viewing angles (Sect.~\ref{subsec:sirocco_results}).

The \textsc{Sirocco} model has an axisymmetric two-component geometry consisting of a polar outflow and an equatorial inflow, first proposed by \citet{Sneppen2026a}, who showed that a bi-conical outflow embedded in an equatorial gas distribution can simultaneously reproduce the broad Balmer profiles and Balmer absorption of the LRDs. Physically, a central ionizing source is surrounded by a gas envelope whose \ion{H}{i} column density decreases from the equatorial plane towards the poles: the highest equatorial columns produce the strongest Balmer breaks, while the lowest inclinations intercept little \ion{H}{i} and resemble the compact blue BLEs studied here (whose broad Balmer wings nonetheless remain shaped by the high column of ionized gas in the envelope). This viewing-angle sequence naturally reproduces the full spectral diversity of the population, from the reddest, most absorbed LRDs to the unobscured compact blue BLEs. However, we note that a suite of spherically symmetric models with (evolutionary-)variations in the column densities would yield qualitatively similar results, as long as a similar density and ionization structure along the line of sight can be captured.

\begin{figure*}
\centering
\includegraphics[width=\textwidth]{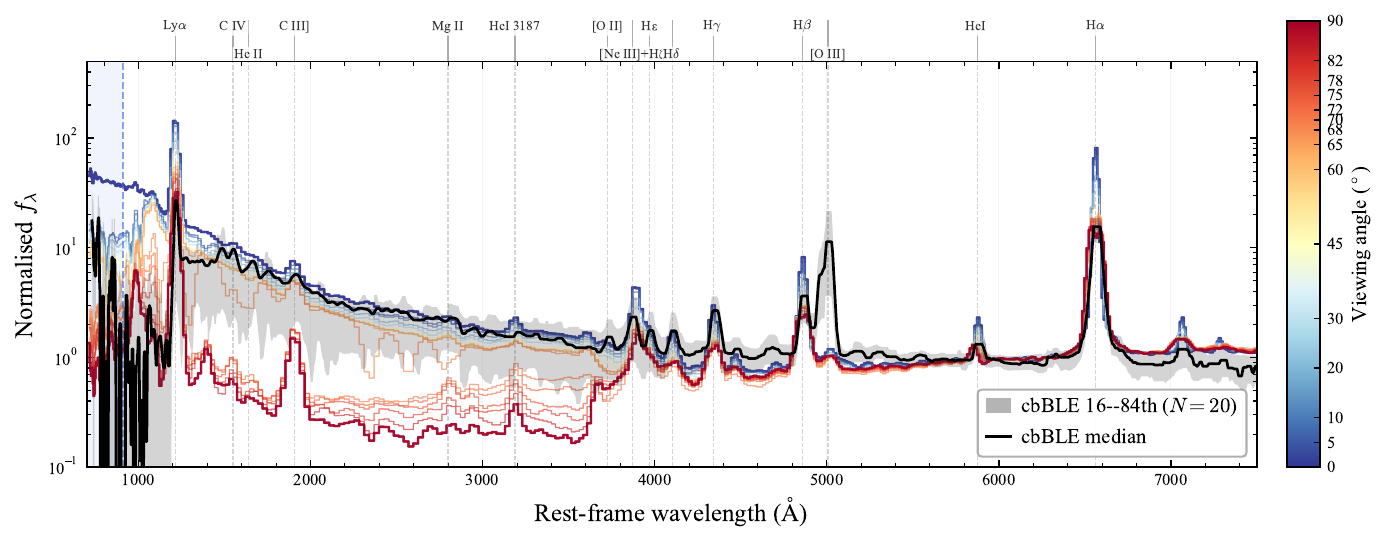}
\caption{Best \textsc{Sirocco} radiative transfer model spectra degraded to NIRSpec/PRISM resolution, shown for all viewing angles from pole-on (blue, $\theta = 0\degr$) to edge-on (red, $\theta = 90\degr$), compared to the mean stacked spectrum of our sample of compact blue broad-line emitters (black), along with the 16th and 84th percentiles. All spectra are normalized to the median flux in the $6200$--$6500$\,\AA\ rest-frame window. The shaded blue region marks the LyC ($\lambda < 912$\,\AA) region. Key features are labeled at the top.}
\label{fig:stack_sirocco}
\end{figure*}

In our best models, the two zones are separated by the boundary angle $\theta_{\rm b}\approx70.5\degr$ (measured from the polar axis), corresponding to a solid-angle ratio $\Omega=\cos\theta_{\rm b}/(1-\cos\theta_{\rm b})=0.5$ between the components; the polar outflow thus subtends approximately one third of the total solid angle, broadly consistent with the fraction of broad-line emitters showing Balmer absorption -- a signature of the equatorial inflow -- among LRD samples \citep[$\approx30-65\%$;][]{Matthee2026, Juodzbalis2026}. The transition between the two zones is smoothed over a $3\degr$ taper to avoid numerical discontinuities at the boundary.

The gas moves radially at $v(\theta)=v_{\rm out}=700$\,km\,s$^{-1}$ in the polar region ($\theta<\theta_{\rm b}$) and inflows at $v(\theta)=\eta\,v_{\rm out}$ with $\eta=0.01$ in the equatorial region ($\theta>\theta_{\rm b}$). For a steady-state wind of mass-loss rate $\dot{M}$ spanning inner and outer radii $R_{\rm in}$ and $R_{\rm out}$, mass conservation fixes the density, 
\begin{equation}
  \rho(r,\theta) = \frac{\dot{M}}{4\pi r^2\, v(\theta)}, \qquad
  v(\theta) = \begin{cases} v_{\rm out}, & \theta < \theta_{\rm b} \\
  \eta\, v_{\rm out}, & \theta > \theta_{\rm b}, \end{cases}
  \label{eq:density}
\end{equation}
which yields the $\rho\propto r^{-2}$ radial dependence of a steady wind and an equatorial-to-polar density enhancement $\rho_{\rm in}/\rho_{\rm out}=1/\eta=100$. We normalize $\dot{M}$ so that the polar sight lines intercept a column density of $\log N_{\rm H}\approx23.5$\,cm$^{-2}$, consistent with the values inferred for the compact blue BLEs \citep{Sneppen2026b}; the equatorial sight lines then reach $\log N_{\rm H}\approx25.5$\,cm$^{-2}$, characteristic of the heavily absorbed red LRDs \citep[e.g.,][]{Matthee2026}. Crucially, the emergent broad line widths are not set by this bulk motion but arise primarily from electron scattering in the dense envelope, with profiles dependent on the electron-scattering optical depth and temperature, the formed determined by the density structure of Eq.~\ref{eq:density} \citep{Sneppen2026a}.

Keplerian rotation $v_\phi=\sqrt{GM_{\rm BH}/r}\,\sin\theta$, where $M_{\rm BH} = 10^7 M_\odot$ is the mass of the central source, is included to provide angular-momentum support in the equatorial gas, though the emergent spectra along polar sight lines are only weakly sensitive to this term. Through Eq.~\ref{eq:density} the radial density slope also governs the ionization structure of the envelope: steeper profiles concentrate the ionized region near the inner boundary, while shallower ones let the ionization front extend outward. Finally, the medium is clumpy, with a volume filling factor $f=0.01$; this raises the local recombination rate by $1/f=100$ relative to a smooth medium of the same mean density, enhancing recombination-line emission while allowing high column densities at lower volume-averaged density.

The central ionizing source is modeled as a blackbody with temperature $T_{\rm BB}=5.0\times10^{4}$\,K and bolometric luminosity $L=10^{44}$\,erg\,s$^{-1}$, typical of LRDs \citep{deGraaff2025b}. We adopt this simple parameterization rather than an AGN power-law spectrum: as discussed in Sects.~\ref{subsec:optical} and \ref{subsec:uv}, the emission-line diagnostics do not unambiguously identify the central engine, and a blackbody provides a physically agnostic but sufficiently hard ionizing continuum to reproduce the observed high-ionization lines without presupposing an accreting black hole. The wind metallicity is set to $Z=0.01\,Z_\odot$ \citep[consistent with][]{Gentile2026}.

The choice of viewing angle for comparison with the compact blue BLE stack follows naturally from this geometric picture. Their lack of Balmer absorption and blue UV continua suggest sight lines dominated by the polar direction, i.e.\ $\theta \ll \theta_{\rm b}$. We therefore compare the stacked spectrum primarily with model spectra extracted at $\theta = 0$--$30\degr$, while using the full range of viewing angles ($\theta = 0\degr$--$90\degr$) to illustrate the diversity of emergent spectral morphologies and the connection between compact blue BLEs and LRDs within a unified framework. The fixed model parameters are summarized in Table~\ref{tab:sirocco_params}.

\subsection{Emergent spectra and viewing-angle dependence}
\label{subsec:sirocco_results}

To compare directly with the observations, we degrade the angle-specific model spectra to the NIRSpec/PRISM resolution ($R \sim 100$) and rescale the flux to the luminosity distance at the median redshift of our sample ($z = 5.3$). The degraded spectra are shown in Fig.~\ref{fig:stack_sirocco}, overlaid on the stacked PRISM spectrum of the compact blue BLE sample. 

\begin{figure}
\centering
\includegraphics[width=0.9\columnwidth]{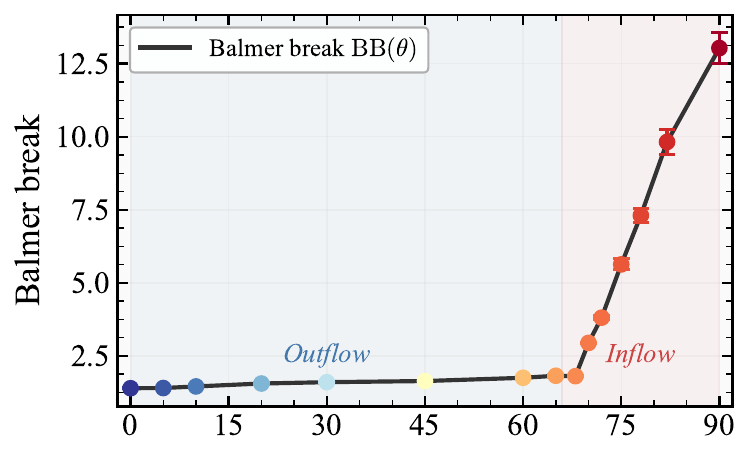}
\includegraphics[width=0.9\columnwidth]{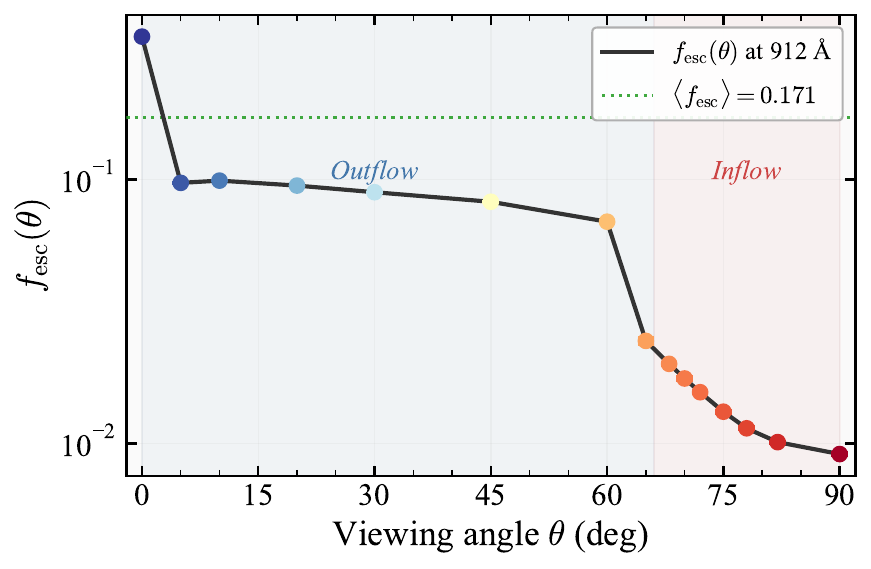}
\caption{Viewing-angle dependence of the Balmer break strength (top) and LyC escape fraction (bottom) for the best-fit \textsc{Sirocco} model. The blue and red shaded regions denote lines of sight through the polar outflow ($\theta \lesssim 70.5\degr$) and equatorial inflow ($\theta \gtrsim 70.5\degr$), respectively. The Balmer break strength rises sharply across the outflow–inflow transition, coincident with the increase in neutral hydrogen column density along equatorial sightlines. This enhanced neutral column renders the gas optically thick to ionizing photons, producing the corresponding decline in the LyC escape fraction, $f_\mathrm{esc}(\theta)$, at $\theta \gtrsim 70\degr$.}
\label{fig:bb_fesc_angle}
\end{figure}

Our stack samples only the weak-break, near-pole-on end of the sequence. To test the model across the full range of Balmer break strengths, we additionally compare it to the LRD stacks of \citet{Matthee2026}, binned by Balmer break strength into four bins of increasing break values. Because the Balmer break is set by the neutral-hydrogen column along the line of sight, which in our geometry increases with inclination, these bins probe progressively more equatorial sight lines. We assign to each bin the model angle that minimizes the reduced $\chi^2$ between the binned stack and the degraded model spectra, computed jointly with our compact blue stack: this yields best-matching angles of $\theta\approx[0,5,68,82,90]$\degr and a total reduced chi-square of $\chi^2_\nu = 12$. The recovered trend between the viewing-angle and the Balmer break strength, measured as discussed in Sect.~\ref{sec:beta_selection}, is shown in Fig.~\ref{fig:bb_fesc_angle}, top panel.

\subsection{Constraints on the ionizing photon production and escape}
\label{subsec:lyc_escape}

We use the \textsc{Sirocco} model spectra to derive $\xi_{\rm ion}$ and $f_{\rm esc}^{\rm LyC}$ as a function of viewing angle. The ionizing photon rate, expressed in units of photons per second, is computed as
\begin{equation*}
Q(\theta) = \int_0^{912\,\text{\AA}} \frac{L_\lambda(\theta)\,\lambda}{hc}\,\mathrm{d}\lambda,
\end{equation*}
where $L_\lambda(\theta) = 4\pi D_{\rm model}^2 F_\lambda(\theta)$ is the angle-specific luminosity at the model distance $D_{\rm model} = 100$\,pc. The total escaping ionizing budget is $Q_{\rm tot} = \sum_\theta Q(\theta)$, and the viewing-angle-dependent escape fraction is defined as
$
f_{\rm esc}^{\rm LyC}(\theta) = Q(\theta)/Q_{\rm int},$ 
where $Q_{\rm int}$ is the intrinsic ionizing photon production rate of the central source prior to radiative transfer. Fig.~\ref{fig:bb_fesc_angle}, bottom panel, shows the viewing-angle-dependent escape fraction, which ranges from $\sim$0.01 to $\sim$0.35: the polar viewing angles have the highest values due to the lower column density along the outflow cone. We assess the robustness of $f_{\rm esc}^{\rm LyC}(\theta)$ by sweeping the upper integration cutoff over $\lambda_{\rm cut} \in [912 \pm 20]$\,\AA\ in 41 steps, propagating the resulting variance as a systematic uncertainty. 

To estimate $\xi_{\rm ion}$, we use the intrinsic spectrum to compute both $Q_{\rm int}$ and the monochromatic UV luminosity $L_\nu(1500\,\text{\AA})$, giving $\xi_{\rm ion} = Q_{\rm int}/L_\nu(1500\,\text{\AA}) [\text{Hz erg}^{-1}].$ The predicted $\log\xi_{\rm ion}$ and the viewing-angle distribution of $f_{\rm esc}^{\rm LyC}$ are discussed in Sect.~\ref{subsec:implications}.

\section{Discussion}
\label{sec:discussion}

\subsection{The nature of compact broad-line emitters: a coherent physical picture}
\label{subsec:nature}

The spectral analysis of Sect.~\ref{sec:spectra} support a physical picture in which compact blue BLEs and LRDs are drawn from the same parent population, with their observed differences reflecting the orientation of the line of sight relative to the axisymmetric gas cocoon as we modeled with \textsc{Sirocco} (Sect.~\ref{sec:sirocco}), rather than an intrinsically distinct physical nature. In line with this picture, \cite{Brazzini2026} report that one of the compact blue BLEs in our sample (GS\_3073) is extremely X-ray weak, and we find the same lack of X-ray emission for the remaining sources in our sample with archival X-ray coverage, mirroring the X-ray weakness reported for LRDs \citep[e.g.,][]{Ananna2024, Setton2024, Yue2024}. Both classes also lack significant photometric variability on timescales of months to years \citep[e.g.,][]{ Ji2025, Zhang2025, Brazzini2026,Liu26}. We now bring these threads together and discuss their quantitative implications.

\subsubsection{Viewing-angle classification of the BLE population}

Of the 99 confirmed broad-line emitters in the parent sample, 20 satisfy the compact blue BLE selection criteria ($\beta_{\rm UV} \leq -1.5$, $\beta_{\rm opt} < 0.5$, and $\mathcal{C}_{\rm UV}, \mathcal{C}_{\rm opt} \leq 1.7$), i.e. $\sim$20\% of the PRISM-selected BLE sample.
The remaining $\sim$80\% are predominantly LRDs (see Fig.~\ref{fig:beta_diagram}), characterized by red optical slopes and elevated Balmer breaks. This apparent dominance of LRDs must, however, be interpreted with care, because the PRISM broad-line recovery rate is a strong function of \ha\ luminosity and line width (Sect.~\ref{sec:ha_selection}). Within the \textsc{Sirocco} geometry, we set the wind boundary to $\theta_{\rm b} \approx 70.5\degr$, to reproduce the blue-to-LRD ratio measured directly from the PRISM sample. Correcting for the luminosity- and line-width-dependent incompleteness of the broad-line selection (Sect.~\ref{sec:ha_selection}), which preferentially removes the fainter blue emitters, raises the intrinsic blue fraction only modestly, from $\sim$26\% to $\sim$32\%, corresponding to $\theta_{\rm b} \approx 67\degr$. This $\sim$3\degr\ shift confirms that our adopted geometry is robust against the selection incompleteness: the completeness correction moves the boundary angle by less than the width of the transition taper and leaves the polar-to-equatorial partition essentially unchanged.
In this physical configuration, the pole-on spectrum ($\theta = 0\degr$) reproduces 8 sources; the intermediate polar spectra at $\theta \approx 5\degr$, $20\degr$, and $30\degr$ match 1, 4, and 3 sources respectively, covering the full range of compact blue BLE spectral types. The remaining 4 sources occupy the transition region at $\theta \approx 45\degr$--$65\degr$ (with a mix of polar and equatorial gas along the sight line), with 1 source at $\theta \approx 60\degr$ and 3 at $\theta \approx 65\degr$, consistent with the borderline cases discussed in Sect.~\ref{sec:final_sample}. 

Based on a small sample of LRDs and LBDs at $2 \lesssim z\lesssim 8 $ identified by JWST in the GOODS fields, \citet{Geris2026} find that LRD models require either clumpy or equatorial geometries, consistent with our model. While they favor a ``dust-obscured LBDs'' interpretation, our \textsc{Sirocco} modeling shows that the Balmer break and spectral sequence are primarily governed by gas column density and electron-scattering geometry rather than variations in dust attenuation \citep{Sneppen2026b, Matthee2026}.

\subsubsection{The compact blue BLEs as an unobscured window onto the central engine}

So far, we have remained agnostic about the nature of the central ionizing source, which could range from an accreting black hole to a hot stellar populations in star clusters. The envelope geometry can naturally explain the broad Balmer lines through scattering and radiative transfer effects, meaning that line widths alone do not uniquely diagnose the underlying engine. Instead, we turn to excitation-sensitive UV emission lines, whose ionization potentials and relative strengths provide a more direct probe of the hardness of the ionizing radiation field.

Rest-frame UV medium-resolution grating spectroscopy ($R \sim 1000$) is available for one source in the sample, GN-16813, from the DIVER program (GO-8018, PI Lin). At this resolution, \ion{C}{iv}~$\lambda\lambda1548,1551$ and \ion{N}{iv]}~$\lambda\lambda1483,1486$ are confirmed as the dominant UV emission features, with rest-frame equivalent widths of $\mathrm{EW}_0 \approx 43$ and $31$\,\AA\ respectively. The grating data also recover \ion{He}{ii}~$\lambda1640$, \ion{O}{iii]}~$\lambda\lambda1661,1666$, \ion{C}{iii]}~$\lambda\lambda1907,1909$, and \ion{N}{iii]}~$\lambda1748$ at $\mathrm{EW}_0 \approx 7$\footnote{broad+narrow components}, $14$, $25$, and $5$\,\AA\ respectively. These lines fall entirely below the PRISM detection threshold (Fig.~\ref{fig:comparison}, Table~\ref{tab:GN16813_lines}, in Appendix \ref{sec:GN-16813uv}).

The \texttt{unite} fit to the \ion{He}{ii}$+$\ion{O}{iii]} window reveals that \ion{He}{ii}~$\lambda1640$ requires a broad component in addition to the narrow nebular emission traced by the \ion{O}{iii]} lines, inconsistent with a purely nebular origin. Broad \ion{He}{ii}~$\lambda1640$ with $\mathrm{FWHM} \gtrsim 1000$\,km\,s$^{-1}$ is a well-established signature of very massive stars \citep[VMS;][]{Graefener2015, Marques-Chaves2024}, whose Wolf–Rayet-type stellar winds drive \ion{He}{ii} into broad emission at low metallicity. This interpretation is supported by recent JWST detections of VMS signatures -- strong \ion{N}{v}~$\lambda1240$ and \ion{C}{iv}~$\lambda1550$ P-Cygni profiles alongside broad \ion{He}{ii}~$\lambda1640$ -- in UV-bright galaxies at $z \sim 8.7$ \citep{Marques-Chaves2026}. The \ion{N}{iv]} detection further indicates nitrogen enhancement \citep{Cameron23, Zhang2026}, consistent with the CNO-processed material exposed by these winds. Taken together, the UV spectrum of GN-16813 is not classically AGN-dominated but suggests a significant contribution from an extremely young ($\lesssim 2$\,Myr), massive stellar population. An additional AGN contribution cannot be excluded, however, given the broad \ha\ emission and the line ratios placing GN-16813 in or near the AGN locus of multiple BPT diagnostics (Fig.~\ref{fig:optical_bpt}); this inconsistent classification across diagnostics prevents firm conclusions on the dominant ionizing mechanism and underscores the need for higher-resolution spectroscopy to disentangle star formation, shocks, and AGN activity.

\lya\ emission is detected in the grating spectrum with a redshifted peak at $v_\mathrm{red} \approx +260$\,km\,s$^{-1}$. This relatively small velocity offset points to a low \ion{H}{i} column density along the line of sight, consistent with efficient \lya, and potentially LyC, escape \citep{Verhamme15}. The absence of low-ionization interstellar absorption lines in the spectrum suggests a low \ion{H}{i} covering fraction, which would favor a high escape fraction of ionizing photons \citep[e.g.,][]{Saldana-Lopez2022}. Mapping the complete rest-UV diagnostics (\ion{C}{iv}, \ion{He}{ii}, \ion{O}{iii}], \ion{C}{iii}], \ion{N}{iv}] and \ion{N}{iii}]) at $R \sim 2700$ is therefore the next step, both to settle the nature of GN-16813 and to extend this analysis to the remaining 19 compact blue BLEs lacking deep grating data.

The \heii\,$\lambda4686$/\hb\ ratio provides a diagnostic of the hardness of the ionizing source that is accessible even at PRISM resolution in the brightest sources. An elevated \heii/\hb, above the value predicted by standard photoionization models for star-forming regions, constitutes direct evidence for a photon field capable of doubly ionizing helium, a signature of AGN accretion discs or extreme stellar populations \citep{Shirazi2012,Perez-Gonzales2026}. In the stacked spectrum of the compact blue BLEs, \ion{He}{ii}~$\lambda4686$ is detected with $\log$(\heii/\hb) $\approx -1.0$. This value exceeds that of typical star-forming galaxies and their high-redshift analogs \citep[$\log$(\heii\,$\lambda$ 4687/\hb) $\sim -1.8$,][]{Bian2020}, and is consistent with -- or slightly harder than -- the stacked value of $\log$(\heii\,$\lambda$ 4687/\hb) $= -1.45 \pm 0.09$ reported for DESI LRDs \citep[]{Lin2026}, which itself is interpreted as intermediate between the star-forming and AGN loci \citep{Shirazi2012}. The systematically harder \heii/\hb\ ratio in the compact blue BLEs relative to LRDs is qualitatively expected in the viewing-angle scenario: at pole-on sight lines, the observer intercepts a smaller column of the dense, partially ionized envelope that reprocesses and softens the emergent spectrum at equatorial viewing angles. The stacked \heii/\hb\ ratio therefore supports the interpretation that compact blue BLEs provide a less-reprocessed view of the same underlying engine responsible for LRDs. 

\begin{figure}
\centering
\includegraphics[width=0.9\columnwidth]{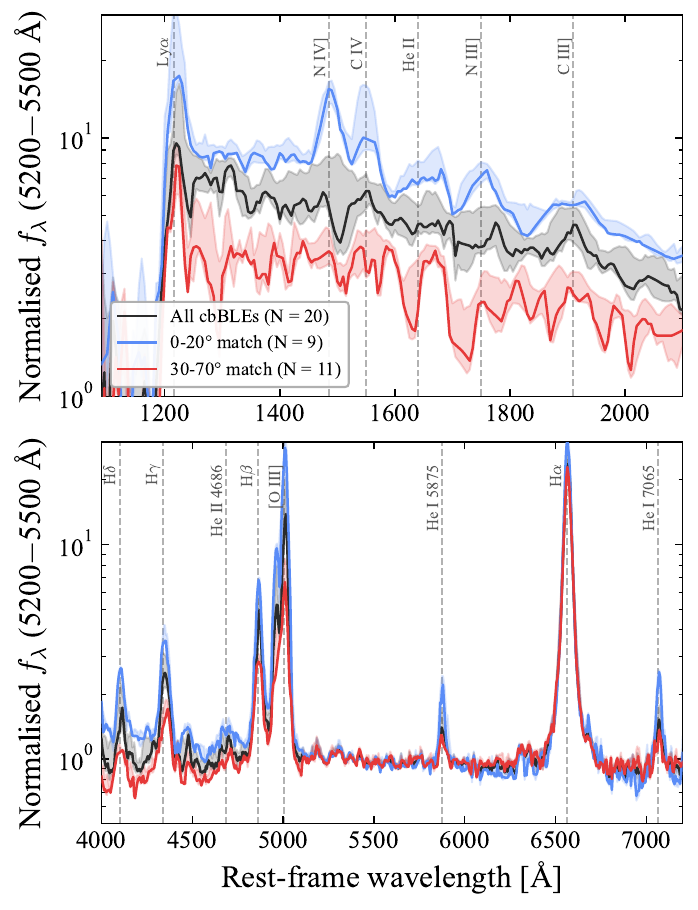}
\caption{Median rest-frame stacked spectra of compact blue BLEs split by assigned viewing angle. The pole-on subsample ($\theta \leq 20\degr$, $N = 9$, blue) and the edge-on subsample ($30 \leq \theta \leq 70\degr$, $N = 11$, red) are compared to the full sample of 20 compact blue BLEs (black). All the stacks are normalized to the median flux in the $5200$--$5500$\,\AA\ rest-frame window; shaded regions show the $1\sigma$ uncertainty on the median. \textit{Top:} rest-frame UV ($1100$--$2100$\,\AA). \textit{Bottom:} rest-frame optical ($4000$--$7200$\,\AA). Key emission lines are indicated by vertical dashed lines. }
\label{fig:stack_angle_split}
\end{figure}

This gradient is directly visible when the sample is split by the viewing-angle assignments of Sect.~\ref{subsec:sirocco_results} (Fig.~\ref{fig:stack_angle_split}). The median stack of the 9 sources best matching the pole-on sight lines ($\theta \leq 20\degr$) displays a systematically bluer and brighter UV continuum, with the detection of \ion{N}{iv]}, \ion{C}{iv]} and \ion{N}{iii]} ($EW_0 = 25.7, 10, 12.5 $\,\AA\ respectively, versus $EW_0 < 3.7, < 3.6, < 4.1 $ for the full sample). These sources also have the largest \lya\ EWs: among the \lya\ emitters in the sample (Sect.~\ref{subsec:lya}), those assigned to $\theta \leq 20\degr$ have a median $\mathrm{EW}_0$(\lya)\, $\approx 152$ \, \AA \ compared to $\approx 42$\, \AA \ for the \lya\ emitters at larger viewing angles. This is precisely the result expected when pole-on sight lines intercept the lowest \ion{H}{i} column densities: both the ionizing continuum and \lya\ photons escape more freely, and these are therefore the sources with the highest $f_{\rm esc}^{\rm LyC}$ in the \textsc{Sirocco} model (Fig.~\ref{fig:bb_fesc_angle}). 

The UV emission lines discussed above are sensitive to hard photons, but can also be powered by hot stellar spectra \citep{Marques-Chaves2026}. A more decisive test of the AGN nature can be provided by [\ion{Ne}{v}]~$\lambda3426$, which requires photons above $97.1$\,eV and is an unambiguous tracer of a very hard radiation field, typically associated with the AGN accretion disc or corona \citep[e.g.,][]{Gilli2010, Mignoli2013}. We searched for [\ion{Ne}{v}]~$\lambda3426$ both in the individual PRISM spectra and in stacked one, finding no significant detection in either. From the stack we place a $3\sigma$ upper limit on the line ratio $\log(\mathrm{Ne53}) \equiv \log([\ion{Ne}{v}]/[\ion{Ne}{iii}]\lambda3869) < -0.5$; combined with the measured $\log([\ion{O}{iii}]\lambda5007/H\beta) \approx 0.6$, this places the compact blue BLEs in the ``composite'' region of the \citet{Cleri2023} diagnostic, and excludes both the traditional supermassive BH accretion-disk AGN locus and the Population~III/intermediate-mass BH region, which require $\log(\mathrm{Ne53}) > -0.3$. This non-detection is physically consistent with the picture emerging from the \heii/\hb\ analysis: the ionizing spectrum  is harder than that of typical star-forming galaxies but softer than that of luminous type-1 AGN or quasars \citep{Wang2025, Ji2025}. Indeed, [\ion{Ne}{v}] has not been reported in any LRD to date \citep{Lin2026,Park26}, and its absence is now understood as a generic property of this class of sources. 

\subsubsection{Average escape fraction across the population}\label{subsec:fesc_bb_calibration}

If the BLE population as a whole -- compact blue BLEs and LRDs alike, excluding the rare quasars with much lower number densities -- is drawn from the same underlying class of objects, then the viewing angle simultaneously determines both the observed spectral type and the $\fesc^{\rm LyC}$.In this framework, the observed distribution of spectral types provides an empirical proxy for the population-averaged $\langle f_{\rm esc}^{\rm LyC}\rangle$, obtained by weighting the escape fraction at each viewing angle by the corresponding fraction of sight lines. The \textsc{Sirocco} model suggests a steep decline of $f_{\rm esc}^{\rm LyC}$ with viewing angle (Fig.~\ref{fig:bb_fesc_angle}): polar sight lines ($\theta\lesssim20\degr$) reach $f_{\rm esc}^{\rm LyC}\approx0.34$, intermediate angles ($\theta\sim45\degr$) $\sim0.02$, and equatorial sight lines ($\theta\gtrsim70\degr$) fall below $10^{-2}$. The solid-angle-weighted average is $\langle f_{\rm esc}^{\rm LyC}\rangle=0.171$, increasing to $\sim0.20$ when weighted by the observed source distribution across angular bins. This same geometry explains the spectral diversity of the population: high-column equatorial sight lines suppress LyC escape and produce strong Balmer breaks, while polar sight lines show weaker breaks and higher escape fractions (Fig.~\ref{fig:bb_fesc_angle}). Since both observables are controlled by the same line-of-sight column density, the Balmer break provides a direct proxy for $f_{\rm esc}^{\rm LyC}$ without requiring individual radiative-transfer modeling.

To quantify this connection, we exploit the full \textsc{Sirocco} model grid rather than the single best-fit model discussed above: we ran 1229 models spanning a range of column densities, wind geometries, and ionizing source properties (Appendix~\ref{appendix:sirocco_grid}). For each model we degrade every viewing-angle spectrum to NIRSpec/PRISM resolution and compare it against the four Balmer-break-binned stacks of \cite{Matthee2026}, which span the dynamic range in the BLE population, assigning to each bin the viewing angle that jointly minimizes the spectral reduced $\chi^2$ and the discrepancy in the \ha\ FWHM. The resulting reduced-$\chi^2$ distribution is strongly bimodal, with a well-separated population of good-fit models at low $\chi^2_\nu$ and a tail of poor fits at $\chi^2_\nu \gg 100$; we therefore retain the 150 models below the natural break in the distribution ($\chi^2_\nu \lesssim 30$). The elevated absolute $\chi^2_\nu$ reflects the very small formal uncertainties of the stacked spectra rather than a poor match to the data (Sect.~\ref{subsec:sirocco_results}), so we use the reduced $\chi^2$ to rank and select models rather than as an absolute goodness-of-fit. While this threshold is not unique, varying it within the low-$\chi^2_\nu$ population does not qualitatively affect our conclusions.

\begin{figure}
\centering
\includegraphics[width=\columnwidth]{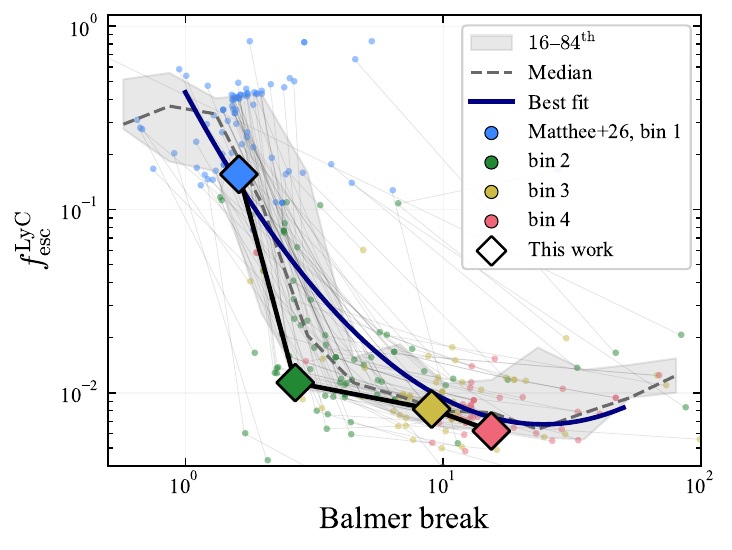}
\caption{LyC escape fraction versus Balmer break strength for the 150 best-fitting \textsc{Sirocco} models, selected from the full grid by minimizing the cumulative reduced-$\chi^2$ across the four Balmer-break-binned stacks of \cite{Matthee2026} and our stack. Each colored point shows one model evaluated at the viewing angle assigned to a given bin (blue~=~bin~1, bluest; green~=~bin~2; yellow~=~bin~3; red~=~bin~4, reddest), with lines connecting the four bins for each model. The gray shaded region marks the $16$th--$84$th percentile range of the population in bins of Balmer break, and the gray dashed line shows the median. The blue curve gives the best-fit second-degree polynomial calibration in log--log space (Eq.~\ref{eq:fesc_bb_calibration}). Our best-fit model from Sect.~\ref{subsec:sirocco_results} is highlighted with black-edged diamonds, falling consistently within the population envelope at every bin.}
\label{fig:fesc_vs_bb_calibration}
\end{figure}

Fig.~\ref{fig:fesc_vs_bb_calibration} shows $f_\mathrm{esc}^\mathrm{LyC}$ as a function of the Balmer break for this selected sample, with each model contributing one point per bin at its assigned viewing angle. Despite spanning a wide range of input physical parameters, the models trace a remarkably consistent trend: $f_\mathrm{esc}^\mathrm{LyC}$ declines steeply with increasing Balmer break before flattening at a residual level of a few $\times 10^{-3}$ for $\mathrm{BB} \gtrsim 10$. Our best model lies within the $16$th--$84$th percentile of the full population at every bin, indicating that it is representative of the broader grid rather than a fine-tuned outlier.

Fitting a second-degree polynomial in log--log space,
\begin{equation}
    \log_{10} f_\mathrm{esc}^\mathrm{LyC} = p_0 + p_1 \log_{10}\mathrm{BB}
    + p_2 \left(\log_{10}\mathrm{BB}\right)^2,
    \label{eq:fesc_bb_calibration}
\end{equation}
we obtain $p_0 = -0.42 \pm 0.07$, $p_1 = -2.2 \pm 0.2$, and $p_2 = 0.66 \pm 0.14$. We use this calibration in Sect.~\ref{subsec:implications} to estimate $f_\mathrm{esc}^\mathrm{LyC}$ for the LRD population from their observed Balmer break strengths.

\subsection{Implications for the contribution to reionization}
\label{subsec:implications}

\begin{figure}
\centering
\includegraphics[width=\columnwidth]{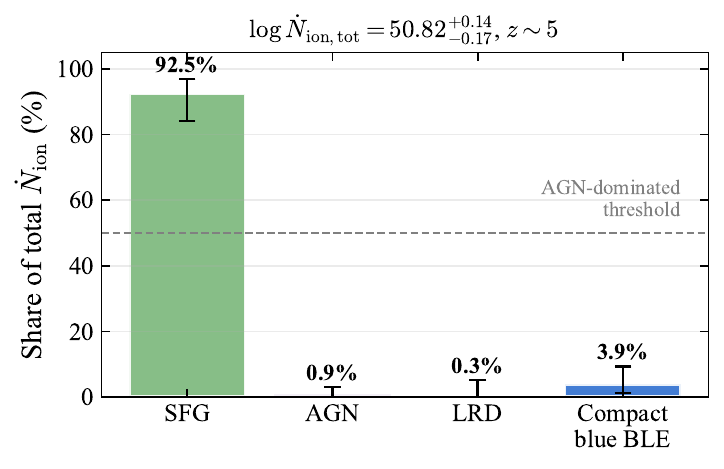}
\caption{Distribution of the total ionizing photon production rate $\dot{N}_\mathrm{ion}$ for the SFG, AGN, LRDs and compact, blue BLEs, following our ``fiducial'' model. Error bars are the 68\% confidence interval from a Monte Carlo propagation of the input uncertainties ($\fesc$, $\xi_{\rm ion}$, $L_{\rm UV}/L_{\rm H\alpha}$, LFs, and population split).}
\label{fig:nion}
\end{figure}

Having established the physical nature of the compact blue BLEs and constrained their ionizing properties through \textsc{Sirocco} modeling, we now quantify their contribution to the reionization photon budget. The total ionizing photon production rate of a source population is given by the luminosity-weighted integral
\begin{equation}
  \dot{N}_{\rm ion} = \int \xion(L_\nu) \, \fesc(L_\nu) \, L_\nu \,
  \phi(L_\nu) \, \mathrm{d}L_\nu,
  \label{eq:nion_integral}
\end{equation}
where $\phi(L_\nu)$ is the UV luminosity function (LF) of the population, and $\xion(L_\nu)$ and $\fesc(L_\nu)$ are allowed to vary with UV luminosity rather than being fixed to a single representative value. This generalization is important because, in our framework, both quantities are governed by the same line-of-sight column density that sets the Balmer break (Sect.~ \ref{subsec:fesc_bb_calibration}): the most UV-luminous, least obscured sight lines preferentially have the lowest Balmer breaks and the highest $\fesc$, while progressively more obscured sight lines have higher Balmer breaks and lower $\fesc$. 

We use the calibration in Eq. \eqref{fig:fesc_vs_bb_calibration} to assign each BLE population a representative $\fesc$ based on its characteristic Balmer break, rather than adopting a single value for all the sources.

Rather than comparing the (incomplete) distributions of the different populations in our total spectroscopic sample (SFG, BLEs -- including classical AGN and LRDs, and the compact blue BLEs), we estimate their contribution to the reionization photon budget from the known luminosity functions. For the star-forming galaxies, we integrate the UV LF of \citet{Bouwens2015} at $z \sim 5$ from $M_{\rm UV} = -24$ down to $M_{\rm UV}=-12$, adopting a canonical $\log\xi_{\rm ion}=25.2$ \citep[e.g.,][]{Llerena2025, Simmonds2024, Mascia24, Prieto-Lyon2023} and average $f_{\rm esc}=0.1$ \citep[e.g.,][]{Mascia24, Simmonds2024, Giovinazzo2026}.

For the BLEs, we integrate the broad-\ha\ LF at $z \sim 4-5.5$ from \cite{Matthee2024} and split it in luminosity: guided by our analysis, we assign emitters brighter than $\log(L_{H\alpha}/\mathrm{erg\,s^{-1}}) = 43.5$ to the classical AGN population and fainter ones to LRDs (66\%) and compact blue BLEs (33\%). We convert \ha\ to an ionizing-photon rate differently for each population: for the AGN, whose broad \ha\ arises in a standard broad-line region, we adopt case-B recombination, $Q = L_{\rm H\alpha}/[1.37\times10^{-12}\,(1-f_{\rm esc})]$; for the LRDs and compact blue BLEs, where the \ha\ is affected by electron scattering in the dense gas, we instead bridge \ha\ to the UV continuum via the measured $L_{\rm UV}/L_{\rm H\alpha}$ ratio, and use $\log\xi_{\rm ion} = 25.4$ estimated from our best model, avoiding a case-B assumption. We adopt escape fractions of $f_{\rm esc}=0.30$ for the AGN \citep[a viewing-angle-weighted average over unobscured and obscured sight lines,][]{D_Amato2020}, $0.2$ for the compact blue BLEs, and $0.01$ for the LRDs.

The resulting escaping ionizing-photon production rate density at $z=5$ is $\log(\dot{N}_{\rm ion}/\mathrm{s^{-1}\,Mpc^{-3}}) =  50.82^{+0.14}_{-0.17}$, consistent with the ionizing emissivity inferred from the \lya\ forest at these redshifts \citep[][]{BeckerBolton2013}. Uncertainties are the $68\%$ confidence intervals from a Monte Carlo propagation of the uncertainties in the adopted $\fesc$, $\xi_{\rm ion}$, $L_{\rm UV}/L_{\rm H\alpha}$ conversions, LF normalizations, and the AGN/non-AGN split. Star-forming galaxies supply the bulk of this budget ($\approx92.5^{+4}_{-8}\%$), while the full broad-line population contributes $\approx5\%$. Within the latter, the compact blue BLEs dominate ($\approx3.9^{+6}_{-2}\%$ of the total), whereas the AGN and the LRDs contribute only $\approx0.9^{+2}_{-0.7}\%$ and $\approx0.3^{+5}_{-0.3}\%$ respectively. As expected by the correlation between Balmer break and $\fesc$, although the LRDs contribute the most to the broad-\ha\ LF, they are negligible reionizers. The broad-line contribution to reionization is therefore carried almost entirely by the minority of low-column sources.  
We consider this the ``fiducial'' model.  Its principal uncertainty is precisely the composition of the broad-line population below our completeness limit, which the luminosity split fixes by construction. Adopting instead an equal split between AGN and non-AGN broad-line emitters, the latter divided between LRDs ($33.5\%$) and compact blue BLEs ($16.5\%$), raises the broad-line share of the ionizing budget to $11.8^{+10.8}_{-6.2}\%$, with the AGN term alone growing from $0.9\%$ to $\sim 8\%$. The two scenarios differ by a factor of $\sim 2.5$ in broad-line emissivity while remaining statistically consistent with one another. Both scenarios fall well short of the $\gtrsim50\%$ required for AGN-dominated reionization \citep{Madau2024}, but a $\sim 5 \%$ contribution is non-negligible, particularly toward the end of reionization, where the ionizing budget is most sensitive to small changes in emissivity \citep{Robertson2015,Finkelstein2019}.
Discriminating between them -- and thus placing the broad-line contribution on a firmer footing -- requires the systematic analysis of higher-resolution grating spectra and deeper surveys, which recover the fainter emitters that the PRISM misses.

A natural consequence of the elevated ionizing output of compact blue BLEs and LRDs is that they may reside within self-generated ionized bubbles, large enough to render the surrounding IGM transparent to their own \lya\ radiation along the line of sight, even at $z\gtrsim7$ \citep{Kokorev23,Tang26}, qualitatively similar to the proximity effect seen around quasars \citep{Eilers25}. This would also help explain the elevated \lya\ emitter fractions reported for this population (Sect.~\ref{subsec:lya}), since efficient \lya\ escape through the IGM typically requires a sufficiently large ionized region around the source \citep[e.g.,][]{Matthee2018,Mason2020}. Following the latter, we estimate the size of the bubbles that could be powered by these sources: at $z = 6$, the bluest sources in our sample ionize bubbles with proper radii of $R_{\rm ion} = 2.15$ pMpc for $\fesc = 0.4$ and $R_{\rm ion} = 1.88$ pMpc for $\fesc = 0.2$, without the contribution from any neighboring sources. At $z = 7$, the same range of escape fractions yields $R_{\rm ion} = 1.61$ pMpc ($\fesc = 0.4$) and 1.41 pMpc ($\fesc = 0.2$),  corresponding to co-moving radii of $\sim$ 15 and 11 cMpc, respectively. Interestingly, these sizes are similar to the scale at which the correlation between the \lya\ transmission excess and galaxy distance is largest during the late stages of reionization \citep{Kashino2026, Jin2026}, i.e. these sizes correspond to the typical regions with excess ionization over the background.

\section{Conclusions}\label{sec:conclusions}

We have presented a systematic census of compact, blue broad-line emitters (BLEs) at $4 \leq z \leq 7$, selected from DJA to isolate sources whose UV continuum and broad emission lines arise from the same compact physical region. By combining rest-frame UV and optical spectroscopy with \textsc{Sirocco} radiative transfer modeling, we address the four questions posed in Sect.~\ref{sec:intro}.
\begin{enumerate}
\item \textit{What fraction of sources exhibit bright, blue UV emission that is predominantly powered by the same region as the broad lines?} Starting from a parent sample of 4145 galaxies with simultaneous \lya-to-\ha\ coverage, we identify 99 broad-line emitters, of which only 20 ($\approx 20$\%) satisfy our combined blue-slope and UV/optical compactness criteria (Sect.~\ref{sec:sample}). 
Although the PRISM broad-line selection is incomplete and biased toward luminous, red sources -- so this  fraction is a lower limit (Sect.~\ref{sec:ha_selection}) -- compact blue BLEs represent a genuine minority of the BLE population, consistent with the geometric expectation from our viewing-angle picture (Sect.~ \ref{subsec:nature}). We emphasize, however, that the adopted thresholds ($\beta_{\rm UV}\leq-1.5$ and $\beta_{\rm opt}<0.5$) are somewhat arbitrary: in the viewing-angle framework, the transition from compact blue BLE to LRD spectral types is continuous rather than discrete, with the observed $\beta_{\rm UV}$ and $\beta_{\rm opt}$ varying smoothly as a function of the line-of-sight column density (Sect.~\ref{subsec:sirocco_results}). 

\item \textit{What are the UV continuum and emission line properties of the blue, compact broad-line emitters as a population?} The compact blue BLEs display blue UV and optical continua, elevated \ha\ equivalent widths, Balmer decrements systematically above the case~B value, strong \ion{He}{i} emission, and a non-detection of [\ion{S}{ii}]~$\lambda\lambda6718,6733$ relative to the SFG population, consistent with a harder ionizing continuum in dense gas (Sect.~\ref{subsec:stack}). Their \lya\ emitter fractions ($X_{\rm Ly\alpha} \approx 50$--$67$\%) are systematically higher than both the SFG and other BLE populations at matched luminosity, pointing to low \ion{H}{i} columns along these sight lines (Sect.~\ref{subsec:lya}).

\item \textit{What spectral diagnostics provide the most effective constraints on the nature and hardness of the ionizing continuum, and can AGN accretion and stellar processes be distinguished using the available data?} High-ionization UV lines (\ion{C}{iv}, \ion{N}{iv]}, \ion{He}{ii}) and an elevated \heii/\hb\ ratio indicate a harder ionizing continuum than typical star-forming galaxies (Sect.~\ref{subsec:uv}). However, neither the rest-frame optical diagnostics (Sect.~\ref{subsec:optical}) nor the UV spectrum of GN-16813 (Sect.~\ref{subsec:nature}) uniquely distinguishes AGN accretion from an extremely young, massive stellar population. The absence of [\ion{Ne}{v}]~$\lambda3426$ throughout the sample further supports an ionizing spectrum harder than normal star formation but softer than luminous Type-1 AGN (Sect.~\ref{subsec:nature}). We therefore adopt a geometry-based interpretation that remains agnostic to the nature of the central engine.

\item \textit{What is the contribution of this population to the reionization photon budget and how sensitive is this estimate to the assumed powering mechanism, surrounding gas geometry, and structure?} \textsc{Sirocco} modeling of an axisymmetric polar-outflow, equatorial-inflow geometry reproduces the spectral diversity from unobscured compact blue BLEs to heavily absorbed LRDs as a function of viewing angle alone (Sect.~\ref{sec:sirocco}). From the full grid of 1229 models, we derive an empirical calibration between the LyC escape fraction and the observable Balmer break strength (Sect.~\ref{subsec:fesc_bb_calibration}, Eq.~\ref{eq:fesc_bb_calibration}). Combining this calibration with the elevated $\xi_{\rm ion}$ predicted by our best-fit model, we find that compact blue BLEs contribute $\approx4$\% of the total ionizing photon budget at $z\sim5$ (Sect.~\ref{subsec:implications}).
While this falls well short of the $\gtrsim 50$\% contribution required for AGN-dominated reionization, it is non-negligible, particularly toward the tail end of reionization where the photon budget is most sensitive to small changes in the emissivity. 
\end{enumerate}
Several caveats apply to these results. Our $f_{\rm esc}^{\rm LyC}$ and $\xi_{\rm ion}$ estimates rely on a single physically motivated wind geometry, calibrated to the stacked spectra; while this geometry self-consistently reproduces the observed spectral diversity across viewing angle, alternative gas distributions or evolutionary histories cannot be excluded with the current data \citep[see][]{Begelman2025, Matthee2026}. The ionizing budget calculation is also sensitive to the faint-end slope and completeness of the BLE UV luminosity function, which remains poorly constrained at $\log L_\nu \lesssim 28$\,erg\,s$^{-1}$\, Hz$^{-1}$
(Sect.~\ref{subsec:implications}). Finally, the nature of the central ionizing source -- whether AGN accretion, extreme massive stars, or some combination of the two -- remains uncertain, and resolving this question will require higher-resolution UV spectroscopy of the kind currently available only for GN-16813 (Sect.~\ref{subsec:uv}).

Our analysis is also limited to $4 \leq z \leq 7$ by the requirement of simultaneous \lya-to-\ha\ coverage with NIRSpec/PRISM (Sect.~\ref{sec:sample}). This is unlikely to represent a physical upper redshift limit for the population: CANUCS-LRD-z8.6 \citep{Tripodi2025}, independently identified in the literature as a classical LRD, in fact satisfies our compact blue BLE selection criteria, and SPURS gdn\_4762\_33609 at $z = 7.18$ \citep{Baccus2025} shows the same absence of \lya\ alongside a rich set of UV emission lines that characterizes our sample (Sect.~\ref{sec:conclusions}). The existence of these sources beyond $z = 7$ suggests that the compact blue BLE population extends to the earliest stages of reionization, and that our results are likely to be representative of the population's ionizing properties across the full EoR.

Finally, at these redshifts LyC is unobservable due to IGM absorption, so the $f_{\rm esc}^{\rm LyC}$--Balmer break calibration of Sect.~\ref{subsec:fesc_bb_calibration} is necessarily model-based rather than empirically calibrated: $f_{\rm esc}^{\rm LyC}$ cannot be directly measured for any source in our sample, compact blue BLE or LRD alike. Testing this calibration therefore requires LRD-like broad-line emitters at $z \lesssim 3$, where the LyC remains observable against the residual IGM opacity. Identifying such a sample, i.e. broad-line emitters spanning the same range of Balmer break strengths as our high-$z$ population, with deep enough UV spectroscopy to either detect or place stringent limits on their LyC emission, would provide a critical empirical check on whether the Balmer break is, as our models predict, a reliable proxy for the ionizing escape fraction at the EoR itself.

\begin{acknowledgements}
SM thanks Nicola Mascia for useful discussions on efficient sample selection.

Funded by the European Union (ERC, AGENTS,  101076224). Views and opinions expressed are however those of the author(s) only and do not necessarily reflect those of the European Union or the European Research Council. Neither the European Union nor the granting authority can be held responsible for them. 

This work is based on observations made with the NASA/ESA/CSA James Webb Space Telescope. The raw data were obtained from the Mikulski Archive for Space Telescopes at the Space Telescope Science Institute, which is operated by the Association of Universities for Research in Astronomy, Inc., under NASA contract NAS 5-03127 for JWST. These observations are associated with programs 1180, 1181, 1208, 1212, 1213, 1215, 1216, 1286, 1345, 1433, 2198, 2561, 2750, 2767, 4106, 4233, 5105, 5224, 5664, 6368, 6585, 8018, and 8204. The authors acknowledge the observing teams that adopted a zero‑exclusive‑access period for several of the programs listed above, making those data public immediately.

The data products presented herein were retrieved from the Dawn JWST Archive (DJA). DJA is an initiative of the Cosmic Dawn Center (DAWN), which is funded by the Danish National Research Foundation under grant DNRF140.

This work was supported by the International Space Science Institute (ISSI) in Bern, through ISSI International Team project \#25-659 ‘Little Red Dots, Big Open Questions’. 

The authors used Claude (Anthropic) to assist with code development for the data analysis pipelines. All AI-assisted content was reviewed, verified, and edited by the authors, who take full responsibility for the accuracy and integrity of this work.

\end{acknowledgements}

\bibliographystyle{bibtex/aa}
\bibliography{bibtex/bib}

\begin{appendix}
\section{Total sample}\label{appendix:totalsample}
\begin{figure*}
\centering
\includegraphics[width=\textwidth]{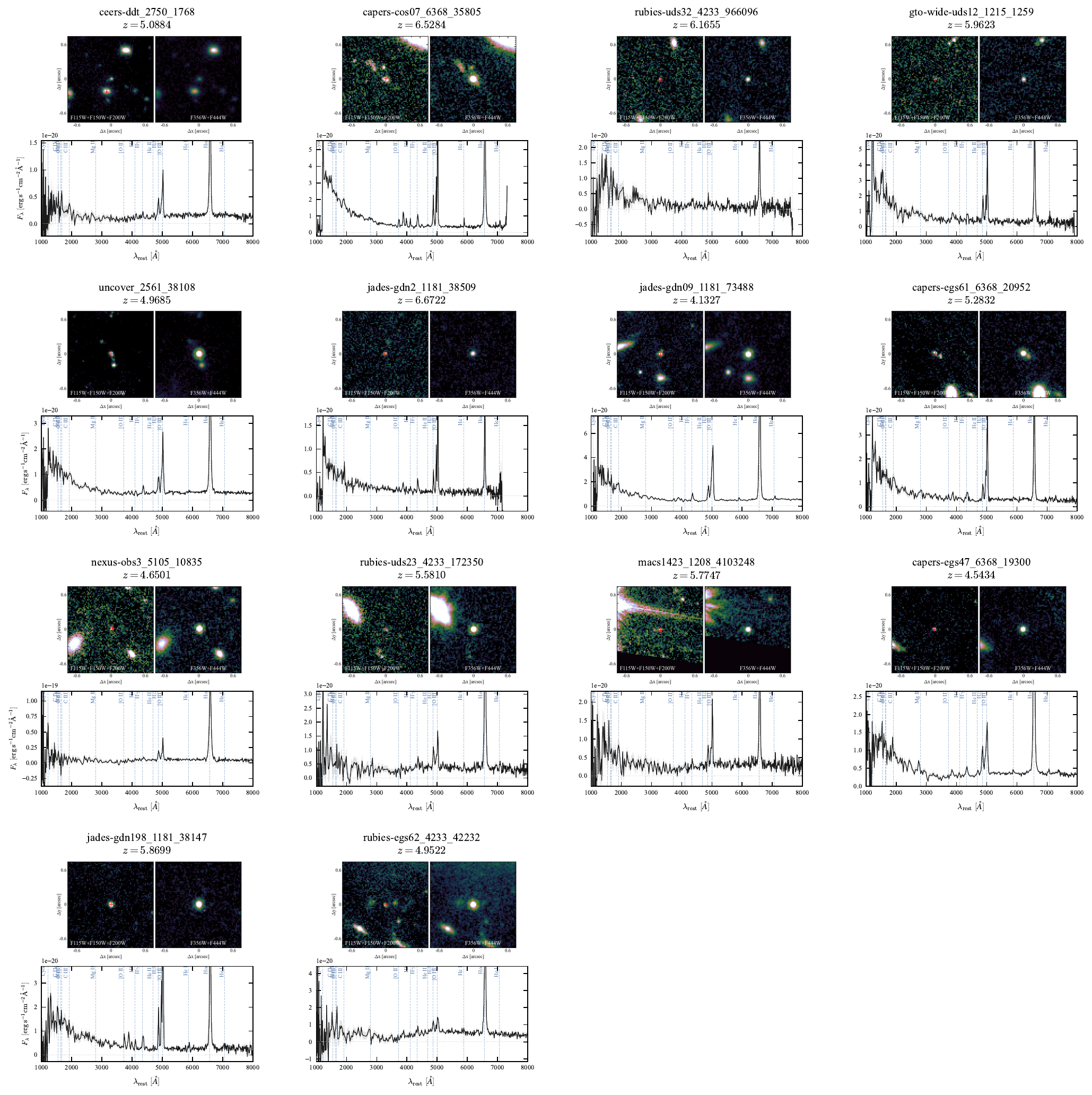}
\caption{NIRCam stamps and NIRSpec/PRISM spectra for the remaining 14 compact blue BLEs. For each source, the upper panel shows NIRCam composites in the rest-frame UV (F115W$+$F150W$+$F200W, left) and optical (F356W$+$F444W, right), with the optical centroid marked by a red cross on the UV stamp. The lower panel shows the PRISM spectrum (black) with the $1\sigma$ uncertainty band (grey). Sources are ordered by increasing $\beta_{\rm UV}$, with the redshift indicated above each stamp.}
\label{fig:stamps_spectra_appendix}
\end{figure*}

Table~\ref{tab:appendix} summarizes the JWST/NIRSpec observing programs and medium- and high-resolution grating coverage available for each of the 20 compact blue BLEs in our sample. Only 12 of the 20 sources have additional coverage with at least one medium- or high-resolution grating, and no source in the sample has the complete combination of all five grating configurations listed. Coverage of the bluest grating, G140M/F100LP, which probes the rest-frame UV at the redshifts of our sample, is available for only 5 sources; of these, GN-16813 is the only source with sufficient depth and wavelength coverage to enable the detailed UV line analysis presented in Sect.~ \ref{subsec:uv}. The remaining 8 sources in the sample have PRISM-only coverage, and their spectral properties are therefore constrained exclusively by the low-resolution analysis of Sect.~\ref{sec:spectra}. Fig.~\ref{fig:stamps_spectra_appendix} shows the NIRCam stamps and NIRSpec/PRISM spectra for the 14 sources not already shown in Fig.~ \ref{fig:stamps_spectra}.

\begin{table*}
\caption{\textbf{Summary of the data available for the 20 compact blue broad-line emitters in 
our sample}.}
\label{tab:appendix}
\centering
\small
\setlength{\tabcolsep}{4pt}
\begin{tabular}{lcccccc}
\hline\hline
Source & Program(s) & G140M/F100LP & G235M/F170LP & G395M/F290LP & G235H/F170LP & G395H/F290LP \\
\hline
jades-gdn09\_1181\_73488  &  1181  & \cmark & \cmark & \cmark & \xmark & \cmark    \\
abell2744-greene\_8204\_69688  & 8204 & \xmark & \cmark & \cmark & \xmark & \xmark\\
capers-egs47\_6368\_19300 & 6368  & \xmark & \xmark & \xmark & \xmark & \xmark    \\
nexus-obs3\_5105\_10835  & 5105 & \xmark & \xmark & \xmark & \xmark & \xmark    \\
rubies-egs62\_4233\_42232 & 4233 & \xmark & \xmark & \cmark & \xmark & \xmark \\
uncover\_2561\_38108  & 2561  & \xmark & \xmark & \xmark & \xmark & \xmark  \\
ceers-ddt\_2750\_1768  & 2750 & \xmark & \xmark & \xmark & \xmark & \xmark  \\
rubies-egs53\_4233\_50052  & 4233 & \xmark & \xmark & \cmark & \xmark & \xmark    \\
capers-egs61\_6368\_20952 & 6368  & \xmark & \xmark & \xmark & \xmark & \xmark    \\
GN-16813  & 5664, 8018  &  \cmark & \xmark & \xmark & \xmark & \cmark        \\
GS\_3073  & 1216 &  \xmark & \xmark & \xmark & \xmark & \cmark      \\
rubies-uds23\_4233\_172350  & 4233 & \xmark & \xmark & \cmark & \xmark & \xmark    \\
jades-gdn11\_1181\_1093  &  1181  & \cmark & \cmark & \cmark & \xmark & \cmark    \\
macs1423\_1208\_4103248  & 1208 & \xmark & \xmark & \xmark & \xmark & \xmark    \\
jades-gdn198\_1181\_38147  &  1181  & \cmark & \cmark & \cmark & \xmark & \cmark    \\
gto-wide-uds12\_1215\_1259  & 1213  & \xmark & \xmark & \xmark & \cmark & \cmark    \\
rubies-uds32\_4233\_966096  & 4233 & \xmark & \xmark & \cmark & \xmark & \xmark   \\
capers-cos07\_6368\_35805 & 6368  & \xmark & \xmark & \xmark & \xmark & \xmark    \\
jades-gdn2\_1181\_38509  &  1181  & \cmark & \cmark & \cmark & \xmark & \cmark    \\
uncover\_2561\_11254 & 2561  & \xmark & \xmark & \xmark & \xmark & \xmark    \\
\hline\hline
\end{tabular}
\end{table*}

\section{SIROCCO}\label{appendix:sirocco_grid}

In addition to the single best-fit model discussed in Sect.~\ref{sec:sirocco}, we explored a broad grid of 1229 \textsc{Sirocco} models to test the robustness of the viewing-angle interpretation and to derive the empirical $f_{\rm esc}^{\rm LyC}$--Balmer break calibration of Sect.~ \ref{subsec:fesc_bb_calibration}. All models share the same underlying two-component geometry described in Sect.~\ref{subsec:sirocco_setup} but span a wide range of central source properties, wind kinematics, geometry, and computational setup. Table~\ref{tab:sirocco_params} lists the parameters set for the best model presented in Sect.~\ref{sec:sirocco}. Table~ \ref{tab:sirocco_ranges} lists the varied parameters across the entire grid. The grid was constructed to broadly cover the parameter space expected for compact, dusty/gas-enshrouded AGN-like sources at sub-parsec to parsec scales, with central black hole masses spanning $10^6$--$10^9$\,M$_\odot$, bolometric luminosities spanning $10^{43}$--$10^{45}$\,erg\,s$^{-1}$, and wind mass-loss rates spanning nearly four orders of magnitude. For each model, spectra were extracted at up to 20 discrete viewing angles spanning $\theta = 0$--$90\degr$ (pole-on to edge-on), enabling us to map the angle dependence of both the Balmer break and $f_{\rm esc}^{\rm LyC}$ self-consistently within each model, as described in Sect.~\ref{subsec:fesc_bb_calibration}.

\begin{table*}
\centering
\caption{Parameters of the best \textsc{Sirocco} model presented in this work.}
\label{tab:sirocco_params}
\begin{tabular}{lll}
\hline\hline
\textbf{Category} & \textbf{Parameter} & \textbf{Value}\\
\hline
\textit{Central source} 
    & Mass & $M_{\rm BH} = 10^7$\,M$_\odot$ \\
    & Bolometric luminosity & $L = 10^{44}$\,erg\,s$^{-1}$ \\
    & Ionizing spectrum & Blackbody, $T_{\rm BB} = 5.0 \times 10^4$\,K \\
    & Schwarzschild radius & $R_{\rm Schw} = 2.95 \times 10^{13}$\,cm \\
    & Disk & None \\
\hline
\textit{Gas envelope}
    & Coordinate system & Polar (axisymmetric) \\
    & Boundary angle & $\theta_{\rm b} = 70.5\degr$ \\
    & Boundary taper & $3\degr$ \\
    & Density profile & $\rho \propto r^{-2}$ [-1, -3]\\
    & Initial temperature & $T_{\rm init} = 10^3$\,K \\
    & Metallicity & $Z = 0.01$\,Z$_\odot$ \\
    & Filling factor & $f = 0.01$ (clumped) \\
\hline
\textit{Kinematics}
    & Polar outflow velocity & $v_{\rm out} = 700$\,km\,s$^{-1}$ \\
    & VRATIO ($\eta$) & $0.01$ \\
    & Density contrast & $\rho_{\rm in}/\rho_{\rm out} = 100$ \\
    & Rotation & Keplerian ($v_\phi = \sqrt{GM/r}\,\sin\theta$) \\
\hline
\end{tabular}
\end{table*}

\begin{table*}
\centering
\caption{Parameters varied across the grid of 1229 \textsc{Sirocco} models.}
\label{tab:sirocco_ranges}
\begin{tabular}{llc}
\hline\hline
\textbf{Category} & \textbf{Parameter} & \textbf{Range}\\
\hline
\textit{Central source}
    & Blackbody temperature & $2.0 \times 10^4$--$1.585 \times 10^5$\,K  \\
    & Bolometric luminosity & $10^{43}$--$10^{45}$\,erg\,s$^{-1}$ \\
    & Black hole mass & $10^6$--$10^9$\,M$_\odot$ \\
    & Radius & $2.95 \times 10^{12}$--$8.857 \times 10^{14}$\,cm \\
\hline
\textit{Stellar wind component}
    & Acceleration exponent & $1$--$2$  \\
    & $\dot{M}$ & $0.011$--$0.702$\,M$_\odot$\,yr$^{-1}$  \\
    & $v_\infty$ & $10^8$--$5 \times 10^8$\,cm\,s$^{-1}$  \\
    & $v_{\rm base}$ & $10^6$--$5 \times 10^7$\,cm\,s$^{-1}$  \\
\hline
\textit{Wind grid and material}
    & Filling factor & $0.001$--$1$  \\
    & $\dot{M}_{\rm wind}$ & $8.67 \times 10^{-4}$--$9.584$\,M$_\odot$\,yr$^{-1}$  \\
    & Metallicity & $0.01$--$1$\,Z$_\odot$  \\
    & $R_{\rm max}$ & $10^{16}$--$10^{19}$\,cm  \\
    & $R_{\rm min}$ & $10^{15}$--$3 \times 10^{16}$\,cm \\
    & Initial temperature & $10^3$--$10^5$\,K \\
\hline
\end{tabular}
\tablefoot{Number of unique discrete values sampled for each parameter is given in parentheses. Each grid point was run with the full set of viewing angles listed in Table~\ref{tab:sirocco_params}, subject to the active viewing-angle configuration for that run.}
\end{table*}

\section{UV lines in GN-16813}\label{sec:GN-16813uv}
\begin{figure}[H]
\centering
\includegraphics[width=\columnwidth]{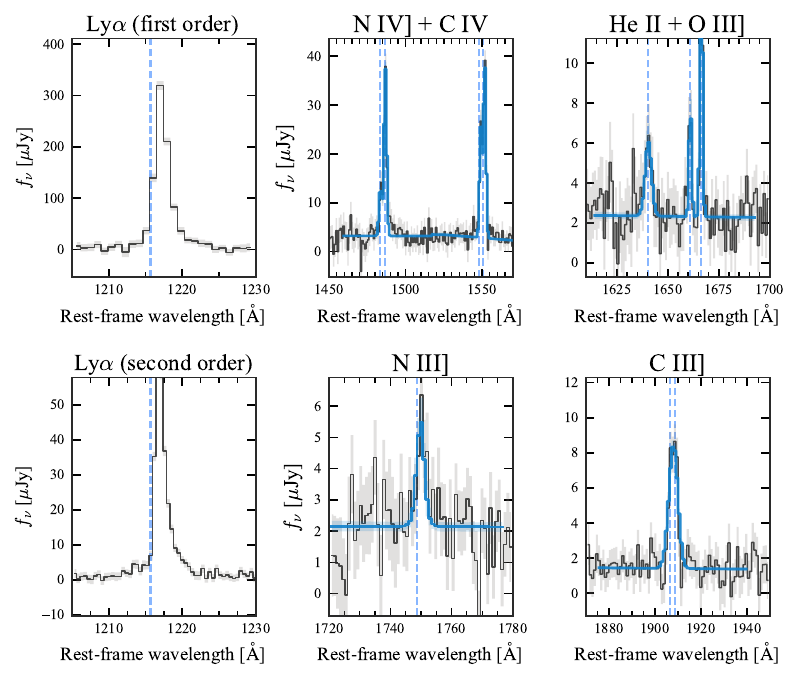}
\caption{Rest-UV spectroscopy of GN-16813. \texttt{unite} best-fit line profiles (solid curves) for the four spectral windows covering \ion{N}{iv]}$+$\ion{C}{iv}, \ion{He}{ii}$+$\ion{O}{iii]}, \ion{N}{iii]}, and \ion{C}{iii]}, overlaid on the continuum-subtracted grating spectrum. On the left panels, first and second order of \lya\ are presented.}
\label{fig:comparison}
\end{figure}

GN-16813 is the only source in our sample with deep medium-resolution NIRSpec grating spectroscopy in the rest-frame UV. We fit \ion{N}{iv]}$+$\ion{C}{iv}, \ion{He}{ii}$+$\ion{O}{iii]}, \ion{N}{iii]}, and \ion{C}{iii]} (Fig.~\ref{fig:comparison}) with \texttt{unite} \citep[see Sec. \ref{subsec:uv},][]{unite_code}. The two panels on the left additionally show the first and second spectral orders of \lya.  
The resulting line fluxes, rest-frame equivalent widths, and diagnostic ratios are listed in Table~\ref{tab:GN16813_lines}. Uncertainties correspond to the 16th--84th percentile range of the MCMC posteriors, and the fluxes are not corrected for extinction; doublet equivalent widths are the summed contribution of both components. 

\begin{table}[H]
\caption{UV emission-line measurements and diagnostic ratios for GN-16813.}
\label{tab:GN16813_lines}
\centering
\small
\setlength{\tabcolsep}{5pt}
\vspace{6pt}

\begin{tabular}{lcc}
\hline\hline
Line & $F$ & $\mathrm{EW}_0$ \\
 & [$10^{-18}$\,erg\,s$^{-1}$\,cm$^{-2}$] & [\AA] \\
\hline
\ion{C}{iv]}$\,\lambda\lambda1548,1551$ & $6.34^{+0.29}_{-0.31}$ & $43^{+5}_{-4}$ \\[2pt]
\ion{He}{ii}$\,\lambda1640$    & $0.98^{+0.23}_{-0.22}$ & $7^{+1}_{-2}$ \\[2pt]
\ion{N}{iv]}$\,\lambda\lambda1483,1486$ & $6.30^{+0.32}_{-0.33}$ & $31^{+3}_{-3}$ \\[2pt]
\ion{N}{iii]}$\,\lambda1748$   & $0.73^{+0.19}_{-0.19}$ & $5^{+2}_{-1}$ \\[2pt]
\ion{O}{iii]}$\,\lambda\lambda1661,1666$ & $2.06^{+0.23}_{-0.20}$ & $14^{+2}_{-2}$ \\[2pt]
\ion{C}{iii]}$\,\lambda\lambda1907,1909$ & $2.24^{+0.16}_{-0.16}$ & $25^{+2}_{-2}$ \\[2pt]
\hline
\end{tabular}

\vspace{6pt}

\begin{tabular}{lcc}
\hline\hline
Ratio & Value \\
\hline
\ion{C}{iv}/\ion{He}{ii}  & $6.47^{+1.48}_{-1.55}$ \\[2pt]
\ion{C}{iii]}/\ion{He}{ii} & $2.29^{+0.54}_{-0.56}$ \\[2pt]
\ion{N}{iv]}/\ion{He}{ii} & $6.43^{+1.48}_{-1.55}$ \\[2pt]
\ion{O}{iii]}/\ion{He}{ii} & $2.10^{+0.53}_{-0.53}$ \\[2pt]
\ion{C}{iv}/\ion{C}{iii]} & $2.83^{+0.24}_{-0.25}$ \\[2pt]
\ion{N}{iv]}/\ion{C}{iii]} & $2.81^{+0.25}_{-0.25}$ \\[2pt]
\ion{N}{iii]}/\ion{C}{iii]} & $0.33^{+0.09}_{-0.09}$ \\[2pt]
\ion{N}{iv]}/\ion{C}{iv}  & $0.99^{+0.07}_{-0.07}$ \\[2pt]
\hline
\end{tabular}

\tablefoot{Fluxes are not corrected for extinction. Doublet EWs are the sum of the two components. Asymmetric uncertainties correspond to the 16th--84th percentile range of the MCMC posteriors.}
\end{table}

\end{appendix}

\end{document}